\documentclass[final,5p,times,twocolumn]{elsarticle}

\usepackage{lineno}
\usepackage{graphicx}
\usepackage{amssymb}
\usepackage{pdflscape}
\usepackage[english]{babel}
\usepackage[dvipsnames,svgnames,x11names]{xcolor}
\usepackage{booktabs,tabularx}
\usepackage{multirow}
\usepackage{hyperref}
\usepackage{url}            
\usepackage{booktabs}       
\usepackage{amsfonts}       
\usepackage{nicefrac}       
\usepackage{microtype}      
\usepackage{amsmath}
\usepackage{commath}
\usepackage[english]{babel}
\usepackage[ruled]{algorithm2e}
\usepackage{caption}
\usepackage{subcaption}
\usepackage{lipsum}
\usepackage{graphicx}
\usepackage{makecell}
\usepackage{longtable}

\usepackage[printonlyused]{acronym}
\usepackage[section]{placeins}
\usepackage[compact]{titlesec} 

\SetEndCharOfAlgoLine{}
\usepackage[leftmargin=6em,rightmargin=12em,indentfirst=false]{quoting}
\usepackage{siunitx}

\usepackage[final]{changes}
\usepackage{float}

\usepackage{lineno}
\usepackage{tikz}
\usepackage[export]{adjustbox}

\usepackage{graphicx}
\usepackage[utf8]{inputenc}
\usepackage[export]{adjustbox}
\usepackage{wrapfig}
\usepackage{dcolumn}

\makeatletter
\newcolumntype{T}[3]{>{\textfont0=\the@{#1}{#2}{#3}}c<{\DC@end}}
\makeatother

\usepackage{pgfplots}
\pgfplotsset{width=10cm,compat=1.9}

\usepackage{array}

\newcolumntype{L}[1]{>{\raggedright\let\newline\\\arraybackslash\hspace{0pt}}m{#1}}
\newcolumntype{C}[1]{>{\centering\let\newline\\\arraybackslash\hspace{0pt}}m{#1}}
\newcolumntype{R}[1]{>{\raggedleft\let\newline\\\arraybackslash\hspace{0pt}}m{#1}}

\usepackage{subfiles}

\usepackage{todonotes}

\RestyleAlgo{ruled}


\journal{Energy and Buildings}
\begin{document}
	
\begin{frontmatter}

\title{InfraRed Investigation in Singapore (IRIS) Observatory: Urban heat island contributors and mitigators analysis using neighborhood-scale thermal imaging}

\author{Miguel Martin$^{1}$, Vasantha Ramani$^{1}$, Clayton Miller$^{2, *}$}

\address{$^{1}$Berkeley Education Alliance for Research in Singapore, Singapore}
\address{$^{2}$College of Design and Engineering, National University of Singapore (NUS), Singapore}

\address{$^*$Corresponding Author: clayton@nus.edus.sg, +65 81602452}

\begin{abstract}
This paper studies heat fluxes from contributors and mitigators of urban heat islands using thermal images and weather data. Thermal images were collected by a rooftop observatory between November 2021 and April 2022. Over the same period, weather stations were operating at several locations on a university campus in Singapore. From collected data, a method was defined to estimate sensible and latent heat fluxes from  building fa\c{c}ades, vegetation, and traffic. Before analyzing heat fluxes using the method, thermal images were calibrated against measurements made with contact surface sensors. Results show that the method can be used to study heat fluxes with a higher temporal resolution at the neighbourhood scale than any other technique using thermal images collected by a satellite. Heat fluxes can also be analyzed over a longer period while considering urban morphology with a higher fidelity than these estimated using most urban climate models. However, as heat fluxes are directly calculated from measurements, the method is not able to forecast the impact of certain elements in the built environment on urban heat islands. In the future, this limitation can be overcome if heat fluxes assessed by the method are used to train and test data driven models.    

\end{abstract}


\begin{keyword}
Urban heat islands \sep Infrared thermography \sep Automatic weather stations \sep Heat balance \sep Building fa\c{c}ades \sep Vegetation \sep Traffic
\end{keyword}

\end{frontmatter}

\section*{Nomenclature}
\small
\begin{tabular}{l l r}
    $\varepsilon$ & Thermal emissivity & 0-1\\
    $\omega$ & Water vapour content & kg$_{water}$/kg$_{air}$\\
    $\phi$ & Relative humidity & 0-1 or \%\\
    $\rho$ & Density & kg/m$^3$\\
    $\sigma$ & Stefan-boltzmann constant (5.67 $\cdot$ 10$^{-8}$) & W/m$^2$-K$^4$\\
    $\tau$ & Solar transmissivity & 0-1\\
    $A$ & Area & m$^2$\\
    $c_p$ & Specific heat & J/kg-K\\
    $c_{veg}$ & Specific heat of vegetation & J/m$^2$-K\\
    $E_{fuel}$ & Fuel consumption & J/m\\
    $e$ & Water vapour pressure & hPa\\
    $h$ & Convective heat transfer coefficient & W/m$^2$-K\\
    $K$ & Shortwave radiation & W/m$^2$\\
    $k$ & Thermal conductivity & W/m-K\\
    $L$ & Longwave radiation & W/m$^2$\\
    $LAI$ & Leaf Area Index & m$^2$/m$^2$\\
    $l$ & Length & m\\
    $N$ & Number of & -\\
    $Q$ & Heat flux & W/m$^2$\\
    $r_a$ & Aerodynamic resistance & s/m\\
    $r_s$ & Stomatal resistance & s/m\\
    $T$ & Temperature & $^o$C or K\\
    $t$ & Time & s\\
    $U$ & Camera output voltage & V\\
    $V$ & Volume & m$^3$\\
    $W_s$ & Wind speed & m/s\\
    $\Delta x$ & Thickness & m\\
\end{tabular}

\section*{List of abbreviations}

\begin{tabular}{l l}
    CFD & Computational Fluid Dynamics\\
    HVAC & Heat, Ventilation, and Air-Conditioning\\
    LST & Land Surface Temperature\\
    MBE & Mean Bias Error\\
    RMSE & Root Mean Square Error\\
    UCM & Urban Canopy Model\\
    UHI & Urban Heat Island\\
\end{tabular}

\normalsize	

\section{Introduction}

Nowadays, more than half of the world's population lives in cities \cite{ritchie2018urbanization}. To accommodate the urban population, buildings and streets have been constructed in large quantities. The expansion of built-up surfaces is a major cause of  Urban Heat Island (UHIs), together with the absence of vegetation and the augmentation of human activity. UHIs are responsible for heat stress in various cities around the world, and thus, it is perceived as a threat to public health in particular during heat wave episodes \cite{campbell2018heatwave}. Due to the importance of this climatic hazard, various methods have been used to study UHIs  \cite{bahi2020review}. While some methods primarily rely on measurements, others have been used to understand causes and effects of UHIs using an urban climate model.  

One of the first methods has used thermal images collected by satellites to observe UHIs \cite{ngie2014assessment, deilami2018urban}. From the thermal images, it is possible to evaluate the Land Surface Temperature (LST) and its difference between urban and rural areas. They can thus provide indications on how hot or cold is the surface of the built environment of a city in comparison to this of its rural surroundings. However, thermal images alone can hardly be used to assess the air temperature difference between urban and rural areas, which is the most common indicator of UHIs. They can also be taken at large time intervals and at a limited scale. As reported in a recent review published by \cite{martin2022infrared}, satellites usually collect thermal images on a daily basis and show the LST at the city scale.

To quantify the air temperature difference between urban and rural areas at a higher temporal resolution and lower scale, other methods have used data collected by a network of weather stations \cite{tan2010urban, wolters2012estimating, warren2016birmingham, meier2017crowdsourcing, konstantinov2018high, rogers2019urban, meng2020impact}. In urban areas, weather stations are commonly installed at the rooftop of buildings or on lamp posts at the street level. Although UHIs can be observed with a higher temporal resolution at the neighbourhood scale from a network of weather stations, they can be studied within a limited number of positions in the city or rural surroundings. To enhance the spatial resolution of observations made from weather stations, some methods have used interpolation techniques \cite{szymanowski2012local, yang2013spatial, bassett2016observations, chapman2017can, liu2017analysis, yadav2018spatial, shaker2019investigating, caluwaerts2020urban, lam2021improvement}. These methods fail in accurately estimating the air temperature between two positions located far from one and the other in the city. The reason is that the air temperature in a city depends on many factors which can hardly be taken into account by an interpolation technique in space.

Instead of interpolating measurements obtained by weather stations, studies have used Computational Fluid Dynamics (CFD), a highly sophisticated modelling technique of urban microclimates, to improve the spatial resolution with which UHIs can be investigated at the neighbourhood scale \cite{toparlar2017review, mirzaei2021cfd}. Based on spatial and temporal discretization methods of Navier-Stokes equations, CFD aims at estimating air motion and temperature at each cell of a three dimensional mesh. Meshes usually consists of a large numbers of cells, and therefore, CFD requires significant computational efforts to assess air motion and temperature at the neighbourhood scale. A consequence of high computational efforts is that CFD can provide information about UHIs and their countermeasures at the neighbourhood scale within a limited time period.

Whether the air temperature is assessed from weather stations or estimated using CFD, none of these methods can be used to determine how significantly certain elements of the built environment contribute to UHIs at the neighbourhood scale. To evaluate the importance of some contributors or mitigators of UHIs in the built environment, studies have mainly used Urban Canopy Models (UCMs) \cite{ching2013perspective, garuma2018review, jandaghian2020comparing}. UCMs are essentially meant to estimate the air temperature and humidity within a street canyon from sensible and latent heat balances. Heat balances are defined from fluxes emitted by building fa\c{c}ades, the street surface, vegetation, Heating, Ventilation, and Air-Conditioning (HVAC) systems, and traffic. The heat fluxes are estimated directly from an empirical formula or indirectly from heat balances. They can be used to explain the reasons why certain elements of the built environment contribute more to UHIs than others. However, when analysing contributors and mitigators of UHIs using UCMs, it is important to remember that this type of urban climate model have a simplified consideration of urban morphology in which buildings are assumed to have similar dimensions and streets equal width.

In the literature, studies have shown that heat fluxes from contributors and mitigators of UHIs can be analyzed considering the urban morphology with a high fidelity to the one observed in reality. One method is to assess heat fluxes from remote sensed data obtained by a satellite together with measurements collected by weather stations. Satellite data are particularly useful for estimating the net-all wave radiation flux \cite{ma2002determination, chrysoulakis2003estimation, bisht2005estimation, wang2005estimation, tang2008estimation, bisht2010estimation, wu2012estimation, qin2020evaluation}. If combined with weather data, they can be used to evaluate convective and/or latent heat fluxes \cite{norman2000surface, french2003surface, liu2012urban, rios2022novel}. The net-all wave radiation flux, the convective heat flux, and the latent heat flux can be balanced to estimate the anthropogenic heat flux and/or the net-heat storage \cite{kato2005analysis, kato2007estimation, xu2008modelling, weng2013assessing, chakraborty2015assessment, chen2017parameterizing}. Despite the complexity of the analysis that can be made on contributors and mitigators of UHIs using methods combining satellite and weather data, it remains that heat fluxes are assessed with a low temporal resolution at a high scale. In addition to this limitation, there is also the fact that only horizontal surfaces can be observed from a satellite.

To assess heat fluxes from both vertical and horizontal surfaces with a higher temporal resolution at the neighborhood scale, it is now possible to collect thermal images from observatories. An observatory consists of an infrared thermal camera installed at the rooftop level. It can also be composed of a pan/tilt unit to collect thermal images at different positions over time.

So far, a few studies have used this modern technology to analyze contributors and mitigators of UHIs at the neighborhood scale. For example, \cite{richters2009analysis} and Morrison et al. \cite{morrison2020atmospheric, morrison2021urban} assessed the longwave radiation emitted by several urban facets. \cite{sham2012verification} estimated the sensible heat transferred by building fa\c{c}ades to the outdoor air. By sensible heat, it is here referred to the total heat transferred by convection and radiation over a period. In other studies, like \cite{hoyano1999analysis}, only considered the heat transferred by convection. Instead of the heat transferred by convection and/or radiation from building fa\c{c}ades, \cite{dobler2021urban} tried to detect sources of anthropogenic heat where HVAC systems are installed at the rooftop of buildings.

These studies show there are three major gaps in the analysis of contributors and mitigators of UHIs using thermal images collected by an observatory:
\begin{itemize}
    \item Firstly, thermal images have focused on building fa\c{c}ades and HVAC systems, which are contributors of UHIs. Consequently, no information is available on mitigators of UHIs, like vegetation, as seen from an observatory.
    \item Secondly, only the sensible heat flux has been assessed from thermal images, and thus, no observation has been made on the net-all wave radiation flux, the latent heat flux, and the net-heat storage.
    \item Thirdly, there is no thermal image collected by an observatory that shows the impact of traffic on the outdoor environment.
\end{itemize}
To fill these gaps, this study aims at accomplishing the following research objectives using thermal images collected by an observatory and measurements obtained by weather stations: 1) Install an observatory from which building fa\c{c}ades, vegetation, and traffic can be observed with a high temporal resolution at the neighborhood scale, 2) Assess all possible heat fluxes from building fa\c{c}ades and vegetation using thermal images collected by the observatory, and 3) Detect traffic from thermal images to estimate their heat releases. 

Results of this study should be of interest to the scientific community and urban planners. The scientific community would be able to see the contribution of certain elements of the built environment to UHIs from a different perspective and temporal resolution at the neighborhood scale as normally reported in the literature. These observations can also motivate investigations on data driven modelling to predict the evolution of UHIs and the efficacy of mitigation strategies. Data driven models, together with thermal images collected from an observatory, could easily be integrated in a digital twin of a city to help urban planners in determining how to make cities more sustainable and resilient towards UHIs. 

The method used to analyze heat fluxes using thermal images collected by an observatory and data obtained by weather stations is described in Section \ref{sec:materials_and_methods}. It comprises the mathematical formulation of heat fluxes, the method used to collect data from an observatory and weather stations, the procedure used to study the sensitivity of the surface temperature assessed from thermal images with respect to some parameters, and the calibration of the surface temperature obtained from thermal images against measurements of contact surface sensors. In Section \ref{sec:results_and_discussion}, the results of this study are detailed and discussed. Finally, conclusions are explained in Section \ref{sec:conclusions}. 

\section{Materials and Methods}
\label{sec:materials_and_methods}

\subsection{Assessment of heat fluxes}
\label{sec:definition_urban_heat_flux}

This section describes the method to analyze contributors and mitigators of UHIs using thermal images collected by an observatory. Contributors here refer to building fa\c{c}ades and traffic, while mitigators corresponds to vegetation. The influence of contributors and mitigators on the outdoor air temperature and humidity within the urban canopy depends on the magnitude their sensible and latent heat fluxes, respectively. To assess the magnitude of heat fluxes from contributors and mitigators of UHIs, it is not sufficient to use only thermal images collected by an observatory. Temperature, relative humidity, wind speed, and solar radiation must be measured by weather stations at the same time. Using thermal images and weather data, it is possible to estimate sensible and latent heat fluxes from building fa\c{c}ades and vegetation directly from their surface temperature or indirectly from an heat balance. Sensible and latent heat fluxes from traffic are calculated using a method to detect cars from thermal images. 

Any built-up surface in an urban area absorbs and emits heat in accordance with the following heat balance:
\begin{equation}
Q^* = Q_H + Q_G + \Delta Q_S
\label{eq:energy_balance_builtup_surface}
\end{equation}
where $Q^*$ is the net-all wave radiation flux; that is, the heat gained or lost by radiation in the short- and longwave range, $Q_H$ the heat transferred from the surface to the outdoor air by convection, that is the convective heat flux, $Q_G$ the heat transfer from the outer to the inner layer of the built-up surface by conduction, and $\Delta Q_S$ the heat stored by the built-up surface (see \ref{sec:appendix_A}). Due to the difficulty in directly evaluating $Q^*$ from Equation \ref{eq:energy_balance_builtup_surface} using thermal images, it is usually recommended to indirectly assess it from $Q_H$, $Q_G$, and $\Delta Q_S$. $Q_H$ and $\Delta Q_S$, however, can directly be estimated from the surface temperature assessed from thermal images ($T_S$) and measurements of one or several weather stations. Given the surface temperature recorded at Position $ij$ by a thermal image ($T_{ij}$), $T_S$ is expressed as:
\begin{equation}
    T_S = \frac{1}{\left| \mathcal{S}\right|}\sum_{ij \in \mathcal{S}} T_{ij}
    \label{eq:surface_temperature}
\end{equation}
where $\mathcal{S}$ is the set of all Positions $ij$ in the thermal image corresponding to the surface $S$. To evaluate $Q_G$, a contact surface sensor needs to be installed inside a building being seen by the observatory.

Equation \ref{eq:energy_balance_builtup_surface} is valid for opaque surfaces; that is surfaces only absorbing and reflecting incoming solar radiation ($K^\downarrow$). Transparent surfaces also transmit a portion $\tau$ of $K^\downarrow$ into buildings. It means that their energy balance is expressed as:
\begin{equation}
Q^* - \tau K^\downarrow = Q_H + Q_G + \Delta Q_S
\label{eq:energy_balance_builtup_surface}
\end{equation}
where $\tau K^\downarrow$ is the transmitted incoming solar radiation through the transparent surface.

In their vegetated urban canopy model, \cite{lee2008vegetated} expressed the heat absorbed and emitted by vegetation as:
\begin{equation}
    Q^* = Q_H + Q_E + \Delta Q_S
    \label{eq:energy_balance_vegetation}
\end{equation}
where $Q_E$ is the latent heat flux produced by evapotranspiration (see \ref{sec:appendix_A}). Similarly to $Q_H$, it can be assessed from thermal images and weather data. The net-heat stored by the vegetation ($\Delta Q_S$) is calculated as:
\begin{equation}
    \Delta Q_S =  c_{veg} \frac{\Delta T_S}{\Delta t}
    \label{eq:heat_storage_vegetation}
\end{equation}
where $c_{veg}$ is the heat capacitance of the vegetation.   

The heat releases from traffic ($Q_{traffic}$), including both sensible and latent heat fluxes, can be expressed as a function of the number of vehicles ($N_v$) crossing a portion of a road using the formula of \cite{grimmond1992suburban}, that is:
\begin{equation}
    Q_{traffic} = \frac{1}{3600 \cdot A_{road}} \left(N_v \cdot l_{road} \cdot E_{fuel} \right)
    \label{eq:traffic_heat_flux}
\end{equation}
The number of cars crossing a portion of the road at an instant $n$ ($N^n_v$) can be detected from thermal images taken at two subsequent times. It means there is a function $f$ such that:
\begin{equation}
    f(|U'^{n+1}_{\mathcal{R}} - U'^{n}_{\mathcal{R}}|) = N^n_v
    \label{eq:function_number_cars}
\end{equation}
where $U'^{n}_{\mathcal{R}} = U^{n}_{\mathcal{R}} - <U^{n}_{\mathcal{R}}>$ is the thermal image within the region $\mathcal{R}$ taken at time $t_0 + n \cdot \Delta t$ and centered over the average $<U^{n}_{\mathcal{R}}>$. Given $f$, the number of vehicles crossing the portion of the road $\mathcal{R}$ over one hour from time $t$ ($N_v$) can be calculated as:
\begin{equation}
    N_v = \sum_{n = (t - t_0)/\Delta t}^{(t + 3600 - t_0)/\Delta t}{N^n_v} = \sum_{n = (t - t_0)/\Delta t}^{(t + 3600 - t_0)/\Delta t}{f(|U'^{n+1}_{\mathcal{R}} - U'^{n}_{\mathcal{R}}|)}
    \label{eq:number_cars}
\end{equation}

\subsection{Rooftop observatory}

From November 2021 to March 2022, an observatory was operating at the rooftop level to collect thermal images of different buildings on a university campus in Singapore. Singapore is a city-state in South-East Asia located near the equator. At this location, a hot and humid climate is experienced over the year. The air temperature varies between 26 and 30 degrees Celsius on average every month. The monthly average humidity is also relatively constant, with variations between 80 and 90 percent. When not obstructed by buildings, the wind most frequently blows at a speed between 1 and 4 meters per second from the South-East direction.

\begin{figure*}[h!]
    \centering
    \includegraphics[width=14cm]{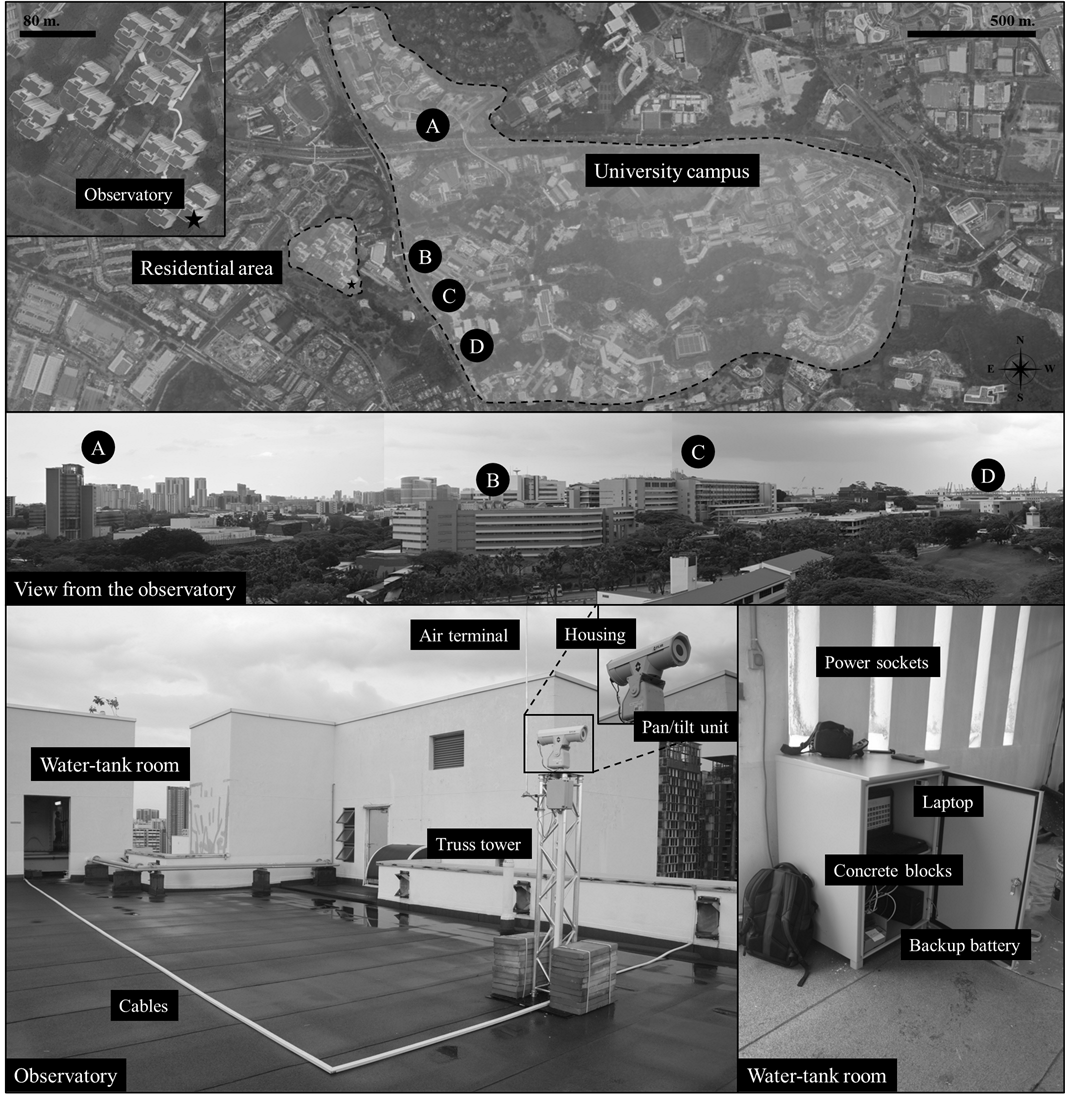}
    \caption{Observatory installed in the university campus in Sinagpore}
    \label{fig:location_view_observatory}
\end{figure*}

The observatory was installed on the rooftop of a 42-meter-tall building located in a residential area, as illustrated in Figure \ref{fig:location_view_observatory}. The residential area is located in front of a university campus consisting of office and educational buildings. Among the buildings, four can be observed from the observatory with a proper resolution. Building A is one of the tallest on the university campus. It is about 68-meter-tall with an important portion of its fa\c{c}ade covered by curtain walls. Closer to the observatory are Buildings B and C, which are both about 27-meter-tall. Their fa\c{c}ade consists of concrete walls and single-pane windows. In addition to concrete walls, the fa\c{c}ade of Building D consists of metal grids installed on a concrete frame. Building D was designed to be net-zero, and its height is around 24 meters. Around buildings A, B, C, and D, it is possible to observe several tropical trees from the observatory. In front of buildings B, C, and D, there is a road with heavy traffic.

On the top of the observatory, there was a housing containing a FLIR A300 (9Hz) thermal camera as described in Table \ref{tab:specifications_a300_thermal_camera}. The housing enables the thermal camera to be protected against heavy rains with IP67 protection. It was fixed on a pan/tilt unit in order to record thermal images at different positions. To avoid any obstacles while recording thermal images, the pan/tilt unit, together with the housing, including the thermal camera, was placed on a 2-meter-high truss tower. This structure is stabilized by concrete blocks and protected against lightning by an air terminal. On the truss tower, two sockets were installed to power up the thermal camera and the direct current motor of the pan/tilt unit from a backup battery located in a water tank room. The backup battery is continuously recharged from the electrical source of the building so as to keep the thermal camera and the pan/tilt unit operating for up to 2 hours in case of power shutdown. The thermal camera and the pan/tilt unit were also connected to a laptop for configuring and checking the collection of thermal images.

\begin{table*}[h!]
    \centering
    \begin{tabular}{l p{6cm} r}
    \hline
    & \textbf{Feature} & \multicolumn{1}{l}{\textbf{Value}} \\
    \hline
     \multirow{7}{*}{\includegraphics[width=2.5cm]{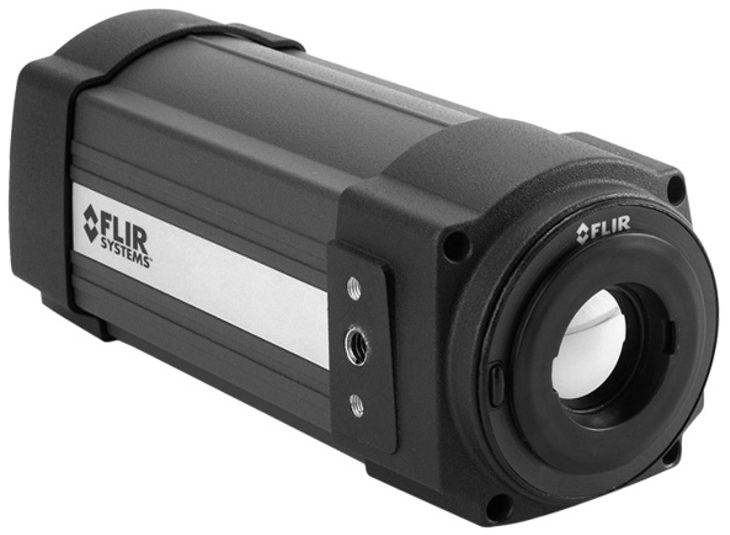}} & Field of view & 25$^o$ (H) x 18.8$^o$ (V)\\
                          & Image resolution & 320 (H) x 240 (V) pixels\\
                          & Accuracy & $\pm$2 $^o$C or 2\% of readings\\
                          & Operating temperature range & -15 … 50 $^o$C\\
                          & Storage temperature range & -20 … 120 $^o$C\\
                          & Spectral range & 7.5 … 13 $\mu$m\\
                          & Image format & 16-bit\\
     \hline
    \end{tabular}
    \caption{Specifications of the FLIR A300 (9Hz) thermal camera.}
    \label{tab:specifications_a300_thermal_camera}
\end{table*}

The collection of thermal images was configured from two separate software. One software was installed on a video encoder to command the pan/tilt unit. From its graphical interface, it is possible to define the positions where the pan/tilt unit must stop to take a thermal image. The moment when a thermal image is taken is controlled by another software installed on the laptop. Thermal images can be saved either in JPEG or FFF file format inside a folder to be specified in the software.

\begin{figure*}[h!]
    \centering
    \includegraphics[width=14cm]{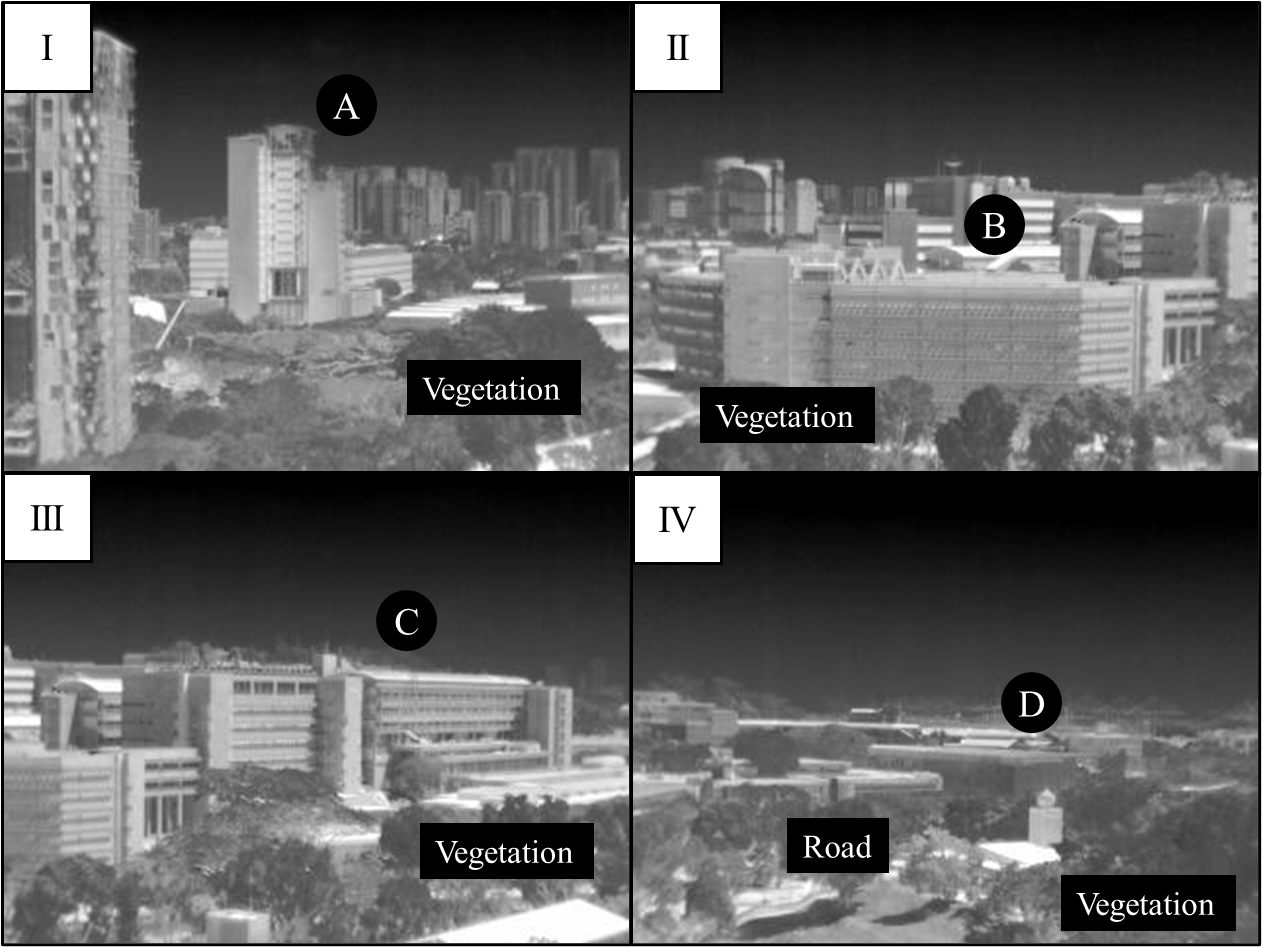}
    \caption{Positions where thermal images were recorded by the observatory.}
    \label{fig:positions_recorded_observatory}
\end{figure*}

The thermal camera and the pan/tilt unit were configured so that images can be taken at four positions, as shown in Figure \ref{fig:positions_recorded_observatory}. Position I is centered on Building A. From this position, it is also possible to observe vegetation consisting of tropical trees mostly. After staying at Position I for a while, the observatory moves to Position II. This position primarily focuses on Building B and its surrounding vegetation. A similar thermal image is taken at Position III but centered on Building D. Finally, the observatory stops at Position IV where various elements can be observed, including Building D, vegetation, and a road. For each position, thermal images are recorded at a rate of one minute approximately. They are stored on a Google Drive repository through a 4G Internet connection installed on the laptop. 

\subsection{Network of automatic weather stations}

\begin{figure*}[h!]
    \centering
    \includegraphics[width=14cm]{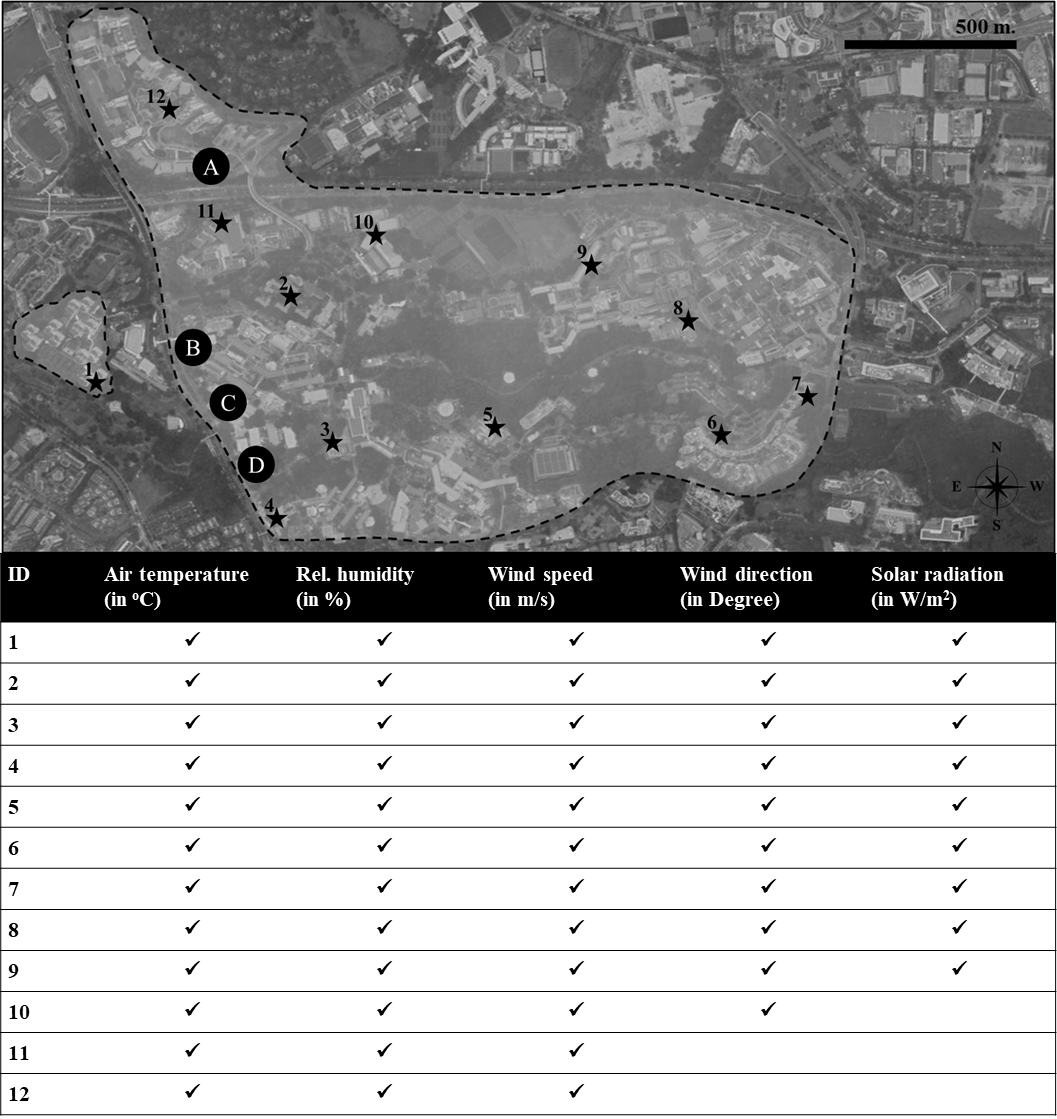}
    \caption{Weather stations installed in a university campus by \cite{yu2020dependence}.}
    \label{fig:awns}
\end{figure*}

Figure \ref{fig:awns} shows the network of weather stations that was used to estimate heat fluxes of built up surfaces and vegetation. The network was deployed by \cite{yu2020dependence} in February 2019. It consists of 12 weather stations measuring the air temperature and relative humidity. All stations, except 12, measure the wind speed and direction. Solar radiation is measured by all stations apart from 11 and 12. Instruments to measure temperature, relative humidity, wind speed/direction, and solar radiation were connected to a data logger to make measurements every 1-minute interval. Their specification is summarized in Table \ref{tab:specification_instruments}.

\begin{table*}[h!]
    \centering
    \begin{tabular}{l r r r r}
    \hline
      Parameter & Temperature  &  Relative humidity & Wind speed & Solar radiation\\
    \hline
       Range  & -40  ... 70 $^o$C & 0 ... 100 \%  & 0  ... 50 m/s & 0  ... 1280 W/m$^2$\\
       Accuracy & $\pm$ 0.21 $^o$C & $\pm$ 2.5 \%  & $\pm$ 1.1 m/s & $\pm$ 10 W/m$^2$\\
       Resolution  & 0.01  $^o$C & 0.05 \% & 0.2 m/s & 1.25 W/m$^2$\\
       \hline
    \end{tabular}
    \caption{Specification of instruments used by \cite{yu2020dependence}}
    \label{tab:specification_instruments}
\end{table*}

As mentioned in Section \ref{sec:definition_urban_heat_flux}, various parameters measured by weather stations need to be used for estimating sensible and latent heat fluxes. For instance, sensible and latent heat fluxes emitted by buildings and vegetation observed at Position I was assessed from the temperature, relative humidity, and wind speed as measured by the Weather Station 12. The latent heat flux emitted by vegetation also depends on the solar radiation, which was defined from measurements of the Weather Station 2. Measurements obtained from the Weather Station 2 were used to evaluate sensible and latent heat fluxes observed at Position II. Heat fluxes observed at Position III and IV were estimated from weather conditions measured by Weather Stations 3 and 5, respectively.  

\subsection{Sensitivity analysis}

Before assessing $T_{ij}$ from the observatory, a sensitivity analysis was conducted on parameters that might affect its variance (see \ref{sec:appendix_B}). A parameters can be a constant or a variable. The contribution of a parameter to the variance of $T_{ij}$ was estimated from the first-order Sobol index \cite{sobol2001global}. The higher the first-order index associated to a parameter is, the more $T_{ij}$ is sensitive to that parameter. However, the first order index does not consider interactions that one parameter might have with others. For this reason, the total Sobol index was also calculated during the sensitivity analysis of $T_{ij}$.

Table \ref{tab:boundaries_variables} illustrate the parameters considered during the sensitivity analysis of $T_{ij}$ and their respective boundaries. According to \cite{osborne2019quantifying}, the thermal emissivity of target object varies between 0.8 and 1.0 in the built environment. In case the emissivity is slightly below 0.8, it was decided to calculate its sensitivity in a range between 0.7 and 1.0. The sky temperature was measured by \cite{miguel2021physically} in Singapore, and varies between 11 and 33 degrees Celsius. Based on weather data recorded by the Meteorological Service of Singapore \cite{mss2022historical}, the outdoor air temperature can change between 19 and 37 degrees Celsius. The air relative humidity is comprised between 30 and 100 percent. Using Google Earth, it was assessed that elements being observed from the thermal camera are located at a distance between 100 and 1000 meters. The surface temperature of the window protecting the infrared camera inside the case of the observatory, as well as its range of values, was measured from a HOBO UX100-014M sensor. A range between 0.8 and 1.0 was considered for transmittance of the window.

\begin{table*}[h!]
    \centering
    \begin{tabular}{p{8cm} r r}
    \hline
    \textbf{Parameter}    & \textbf{Min. value} & \textbf{Max. value} \\
    \hline
     Thermal emissivity (0-1) & 0.7 & 1.0\\
     Sky temperature (in $^o$C) & 11.0 & 33.0\\
     Air temperature (in $^o$C) & 19.0 & 37.0\\
     Air relative humidity (in \%) & 30.0 & 100.0\\
     Distance object (in m) & 100.0 & 1000.0\\
     Window temperature (in $^o$C) & 20.0 & 60.0\\
     Window transmittance (0-1) & 0.8 & 1.0\\
     \hline
    \end{tabular}
    \caption{Boundaries of each variable considered during the sensitivity analysis of  $T_{ij}$.}
    \label{tab:boundaries_variables}
\end{table*}

Given the boundaries described in Table \ref{tab:boundaries_variables}, the first order and total Sobol indices were calculated using the Satteli sampler \cite{saltelli2010variance} on thermal images taken at the four positions during a sunny day in Singapore (i.e., November 17, 2021). It means that the variation of $T_{ij}$ with respect to a given parameter was estimated from a number $R$ of samples, such that:
\begin{equation}
    R = N \cdot (2D + 2)
\end{equation}
where $N = 2^{12}$ and $D=7$. For each sample, $T_{ij}$ was calculated at different times of the day and positions of the thermal camera. The mean indices over the four positions were reported as the sensitivity of $T_{ij}$ with respect to each parameter at different times. 

\subsection{Calibration in an outdoor environment}
\label{sec:calibration}

\begin{figure*}[h!]
    \centering
    \includegraphics[width=14cm]{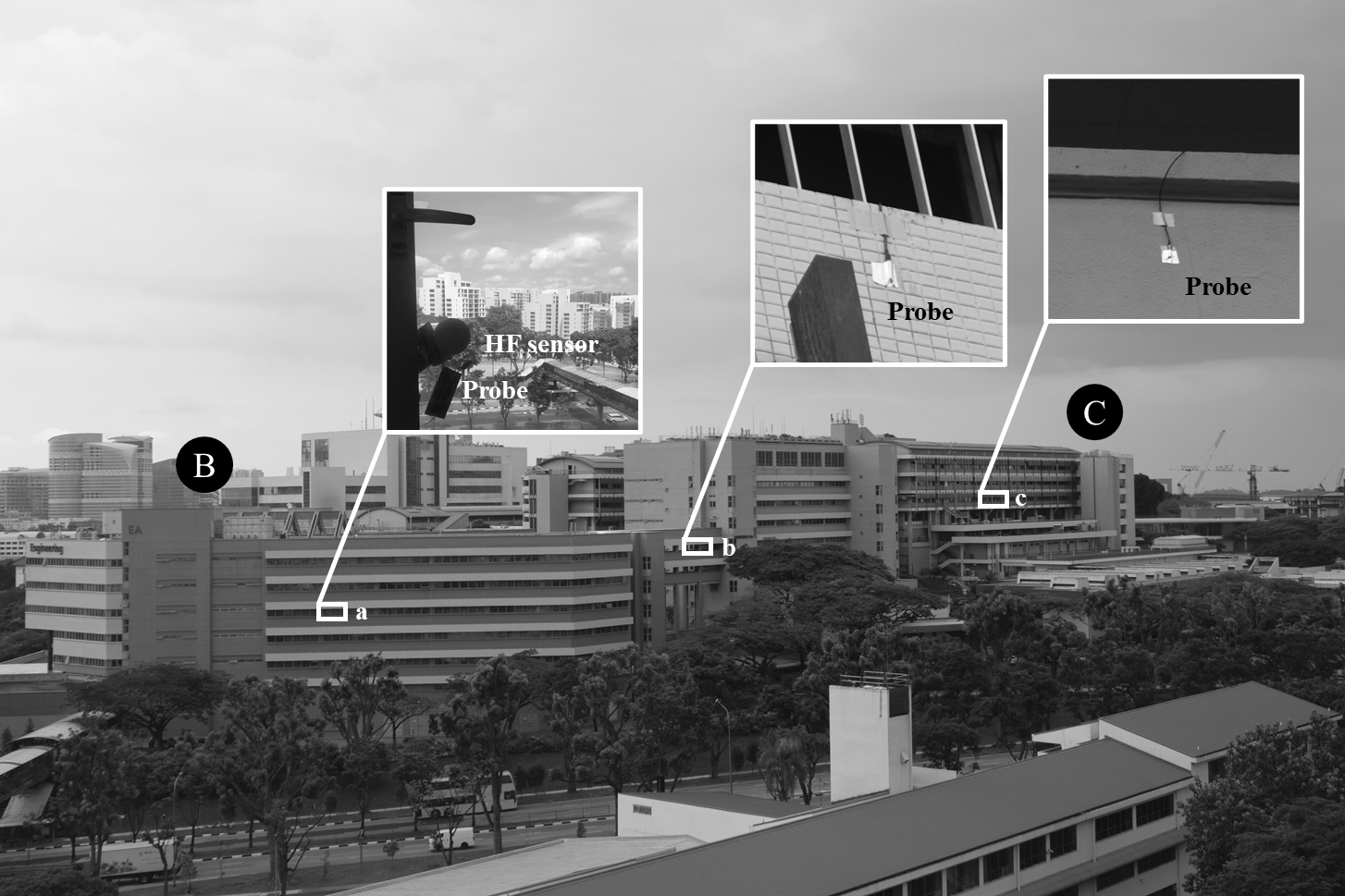}
    \caption{Positions where sensors were installed to measure the surface temperature and calibrate thermal images.}
    \label{fig:calibration}
\end{figure*}

The surface temperature provided by the FLIR A300 camera was calibrated with respect to measurements collected with contact surface sensors. Contact surface sensors were installed at three different positions, as shown in Figure \ref{fig:calibration}. At Position A, one Heat Flux (HF) sensor and one temperature probe were placed at the surface of a window in Building B between December 2021 and March 2022. Both the HF sensors and the temperature probe were connected to a Hioki data logger. With this data logger, the surface temperature was measured every 3 seconds. Another type of datalogger, called UbiBot, was used to connect temperature two and three probes at Positions B and C, respectively. At these positions, the surface temperature was measured during April 2022 at a rate of one minute on concrete walls. During this period, the infrared camera was stopped at Position III and collected thermal images at a higher rate of one per 30 seconds. This choice was made to increase the number of estimates to compare with measurements over a shorter period.

The calibration of the FLIR A300 camera was manually performed by adjusting the parameters of Equation \ref{eqn:surface_temperature_from_output_voltage} till reaching a satisfactory agreement between estimates and measurements of the surface temperature. The agreement was evaluated with respect to the Root Mean Square Error (RMSE) and Mean Bias Error (MBE), which are expressed as:
\begin{eqnarray}
RMSE & = & \sqrt{\frac{1}{N}\sum_{n=1}^{N}{\left(T_S^n -  T_{S,m}^n\right)^2}}\\
MBE & = & \frac{1}{N}\sum_{n=1}^{N}{\left(T_S^n -  T_{S,m}^n\right)}
\end{eqnarray}
where $T_S^n$ is the surface temperature as evaluated from thermal images of the FLIR A300 camera at time $t = t_0 + n \cdot \Delta t$, and $T_{S,m}^n$ the surface temperature measured by contact surface sensors. The objective of the calibration was to reach an MBE as close as possible to 0 with the lowest RMSE. The RMSE and MBE were calculated with different $\Delta t$ at Positions A-C. At Position A, the RMSE and MBE were evaluated with $\Delta t$ equals 30 minutes. A smaller $\Delta t$ of 5 minutes was used for computing the RMSE and MBE at Positions B and C. The reason is that measurements at Positions B and C were collected over a shorter period than at Position A.    

\subsection{Analysis of heat fluxes using thermal images}

From thermal images collected by the observatory, it was possible to analyze heat fluxes within the urban canopy. It includes heat fluxes from fa\c{c}ades of buildings A, B, C, and D, vegetation at Positions I, II, III, and IV, and the traffic on the road observed at Position IV. Among these elements, some can be important contributors of UHIs while others might act as mitigators.

\begin{figure*}[h!]
    \centering
    \includegraphics[width=14cm]{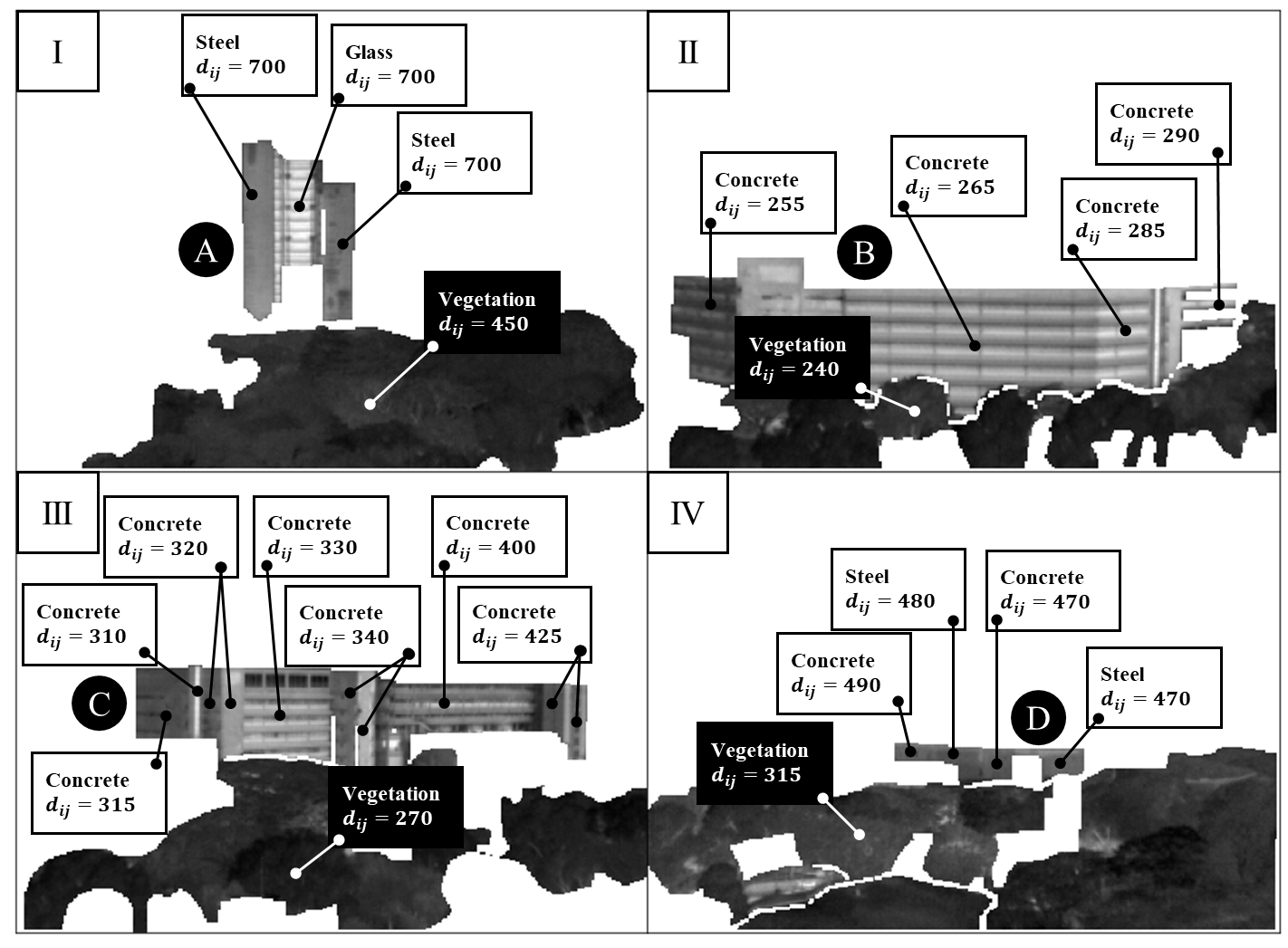}
    \caption{Built-up surfaces and green areas that were considered to analyze urban heat fluxes from the observatory.}
    \label{fig:built_up_green_areas}
\end{figure*}

Building fa\c{c}ades can considerably contribute to UHIs due to the heat they emit during the day and at night. The heat emitted by building fa\c{c}ades, as well as by any built-up surface, depends on their orientation and material properties. For this reason, heat fluxes from fa\c{c}ades of buildings A, B, C, and D were observed from different portions as shown in Figure \ref{fig:built_up_green_areas}. The fa\c{c}ade of Building A consists of two steel walls and one glass wall, whose heat fluxes were assessed separately. Similarly, different portions were considered to observe heat fluxes from fa\c{c}ades of buildings B, C, and D. Portions of buildings B and C are essentially walls consisting of the same material, that is, concrete, but with different orientations. Both steel and concrete walls can be observed on fa\c{c}ades of the building D. Table \ref{tab:thermal_emissivity} shows the thermal properties of building fa\c{c}ades and the vegetation, which were defined from the literature. The distance from the observatory ($d_{ij}$) was measured from Google Earth. The thickness ($\Delta x$) of each built-up surface was assumed to be around 30 centimeters.   

Figure \ref{fig:built_up_green_areas} also shows the different portions of vegetation that were selected in thermal images to assess heat fluxes in the urban canopy. In contrast with building fa\c{c}ades, the vegetation is normally considered a mitigator of UHIs. The reason is that the sensible heat emitted by the vegetation is normally lower than this released by built-up surfaces or any anthropogenic heat source. A considerable portion of the total heat released by vegetation is latent. It implies that vegetation affects outdoor air humidity more than temperature.

 \begin{table*}[h!]
    \centering
 \begin{tabular}{l r r r r r p{1cm}}

    \hline
    \textbf{Portion}    & \textbf{Emissivity} & \textbf{Density} & \textbf{Specific heat} & \textbf{Thickness} & \textbf{Thermal conductivity} &  \textbf{Reference} \\
    & \textbf{(0-1)} & \textbf{(kg/m$^3$)} & \textbf{(J/kg-K)} & \textbf{(cm)} & \textbf{(W/m-K)} & \\
    \hline
     Steel  & 0.88 & 7940 & 507 & 10 & - & \cite{miguel2021physically}\\
                  & & & & & & \cite{meneghetti2013synthesis}\\
     Glass  & 0.93 & 2500 & 836 & 2.5 & 0.974 & \cite{raman1982thermal}\\
                 & & & & & & \cite{ritland1954density}\\
                 & & & & & & \cite{sharp1951effect}\\
                 & & & & & & \cite{wang2020optically}\\
     Concrete  & 0.90 & 2400 & 1180 & 30 & 0.62 & \cite{miguel2021physically}\\
                    & & & & & & \cite{iffat2015relation}\\
                    & & & & & & \cite{de1995specific}\\
                    & & & & & & \cite{kim2003experimental}\\
    \hline
    \textbf{Portion}    & \textbf{Emissivity} & \textbf{LAI} & & & &  \textbf{Reference}\\
    & \textbf{(0-1)} & \textbf{(m$^2$/m$^2$)} & & & & \\
    \hline
     Vegetation & 0.98 & 4 & & & & \cite{olioso1995simulating}\\
     (Tropical trees)& & & & & & \cite{ganguly2008generating}\\
     \hline
    \end{tabular}
	\caption{Thermal properties of building fa\c{c}ades and vegetation observed from the observatory at Positions I, II, III, and IV.}
    \label{tab:thermal_emissivity}
\end{table*}

\begin{figure*}[h!]
    \centering
    \includegraphics[width=13cm]{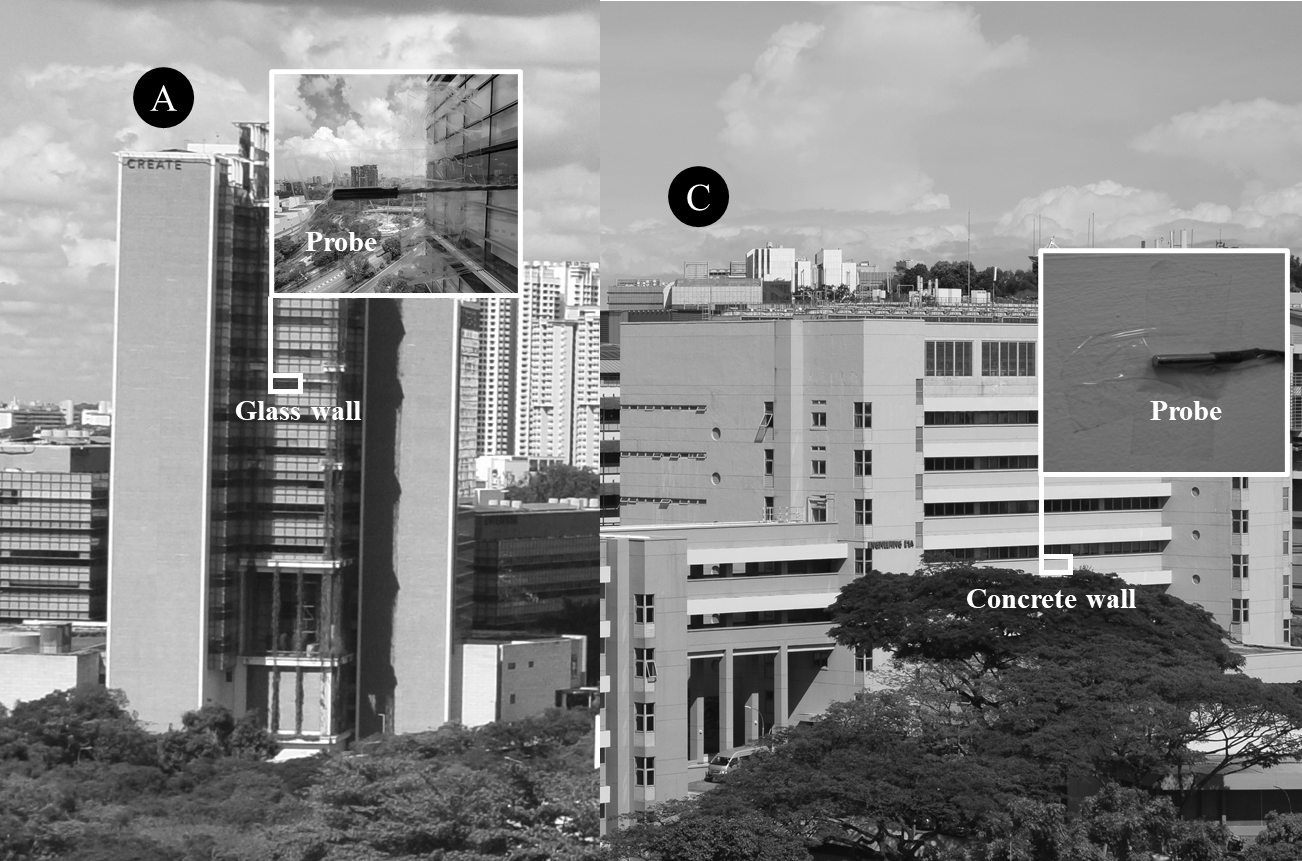}
    \caption{Positions where $Q_G$ was measured with contact surface sensors.}
    \label{fig:indoor_sensors}
\end{figure*}

One heat flux, in particular, had to be measured with additional sensors, the conductive heat flux ($Q_G$). As illustrated in Figure \ref{fig:indoor_sensors}, the conductive heat flux was measured at two different positions: one on the fa\c{c}ade of Building A and another on the fa\c{c}ade of Building C. On the fa\c{c}ade of Building A, the interior surface temperature was measured with a temperature probe connected to a Hioki datalogger, while a UbiBot datalogger was used for measurements on the fa\c{c}ade of Building C. The reason for choosing these two positions was to compare the full energy balance on glass and concrete walls. The conductive heat flux through a glass wall should be higher than this measured on a concrete wall, and thus, it has different implications on the building energy consumption and UHIs.

\begin{figure}[h!]
    \centering
    \includegraphics[width=7.5cm]{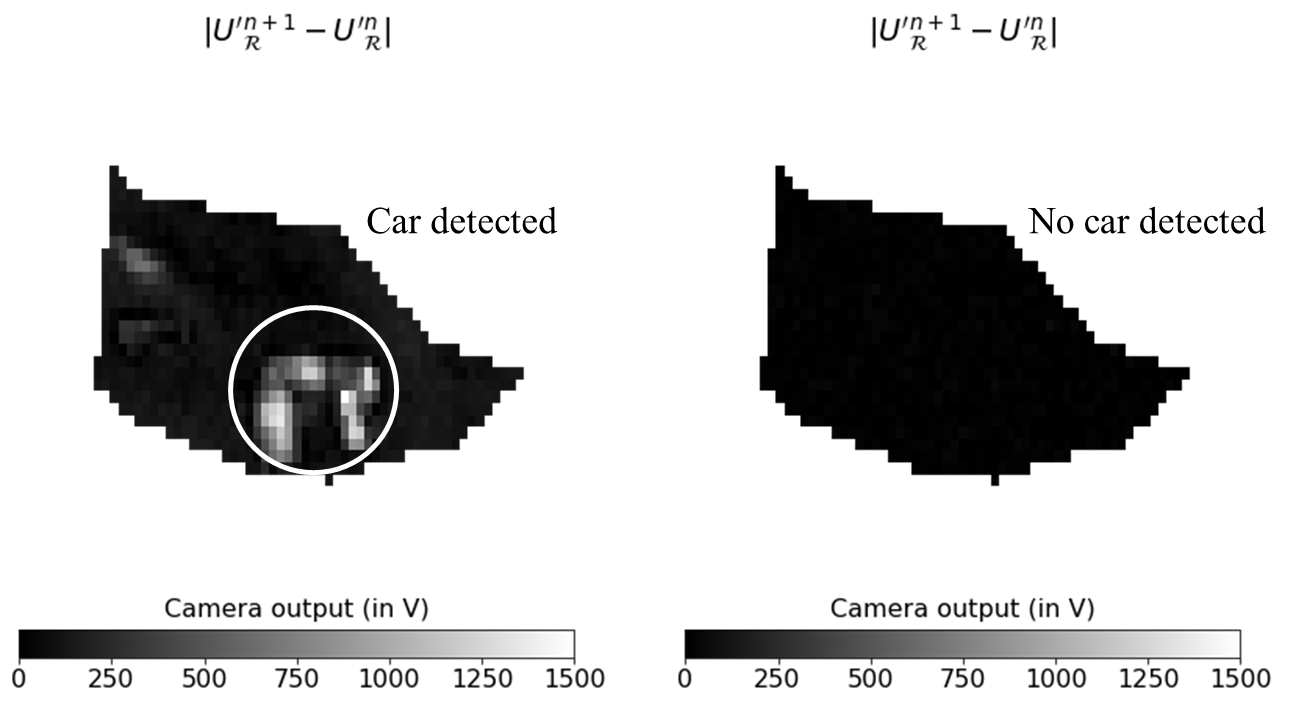}
    \caption{Car detection in a portion of a road at Position IV.}
    \label{fig:car_detection}
\end{figure}

As Illustrated in Figure \ref{fig:car_detection}, a portion of a road at Position IV was selected in thermal images to evaluate the anthropogenic heat flux from traffic. Over this portion, it was assumed that only one car could be detected between two thermal images $U^{n+1}$ and $U^n$. One car is detected if the peak value of $|U'^{n+1}_{\mathcal{R}} - U'^{n}_{\mathcal{R}}|$ is higher than a certain threshold ($\tau$). It means that the function $f$ in Equation \ref{eq:function_number_cars} is formulated as:
\begin{equation}
    f(|U'^{n+1}_{\mathcal{R}} - U'^{n}_{\mathcal{R}}|) = \begin{cases}
    1 & \text{if $\max |U'^{n+1}_{\mathcal{R}} - U'^{n}_{\mathcal{R}}| \geq \tau$} \\
    0 & \text{otherwise}
    \end{cases}
\end{equation}
The value of $\tau$ corresponds to the highest magnitude of the noise that is contained in a thermal image. Regarding the portion of the road at Position IV, the highest magnitude of the noise, that is $\tau$, was approximated as 150 Volts based on observations made on various thermal images collected by the observatory.

\section{Results and discussions}
\label{sec:results_and_discussion}

\subsection{Sensitivity analysis and calibration}

\begin{figure*}[h!]
    \centering
    \includegraphics[width=12cm]{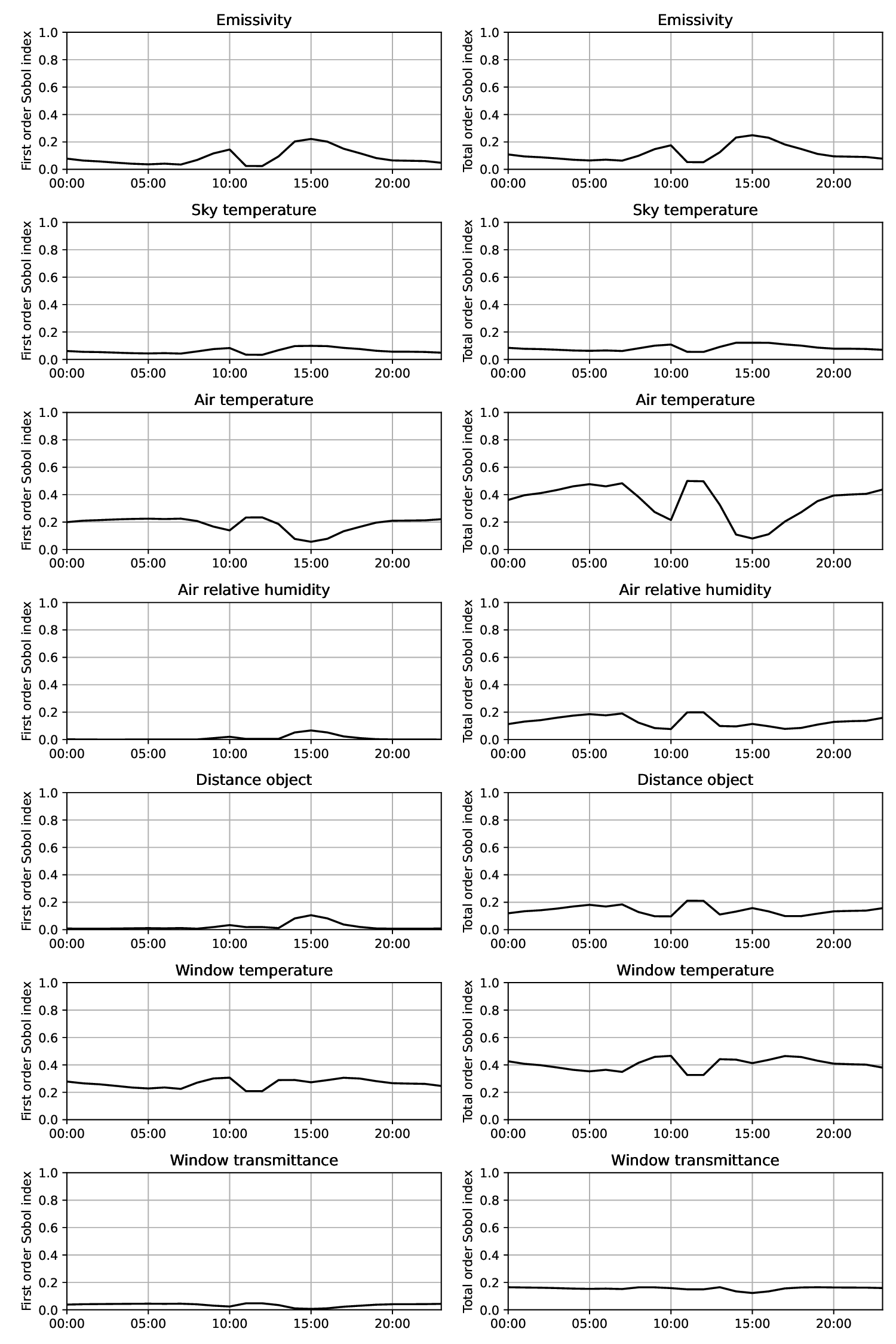}
    \caption{First order and total Sobol indexes of input variables to obtain $T_{ij}$.}
    \label{fig:sensitivity}
\end{figure*}

Figure \ref{fig:sensitivity} shows the sensitivity of temperature values assessed from thermal images ($T_{ij}$) to certain parameters, including the thermal emissivity of the target object, the sky temperature, the air temperature, the air relative humidity, the distance of the target object, the window temperature of the observatory, and its transmittance (see \ref{sec:appendix_B}). It appears that values of $T_{ij}$ are more sensitive to the emissivity, the sky temperature, the window temperature, and the window transmittance around 10 am and 3 pm than the rest of the day. At the observation site, 10 am, and 3 pm corresponds to instants when building fa\c{c}ades are exposed to intense solar radiations. Consequently, it is implied that the solar exposure of building fa\c{c}ades highly affects values of $T_{ij}$ collected by the observatory. In contrast, $T_{ij}$ seems to be more sensitive to the air temperature when building fa\c{c}ades are not exposed to intense solar radiation. The air relative humidity, the distance from the target object, and the window transmittance look to cause minor variations on $T_{ij}$ in comparison to other parameters.  

Due to the sensitivity of $T_{i,j}$ to some parameters, it was important to validate or calibrate its assessed values from the observatory with respect to some other measurements. As explained in Section \ref{sec:calibration}, measurements of the surface temperature were taken with contact surface sensors at three different positions (A-C). The RMSE and MBE obtained after calibration at these three different positions are shown in Table \ref{tab:calibration_a300_thermal_camera}. At the three positions, it was possible to achieve an RMSE below 2 degrees Celsius and an MBE below $\pm$ 1 degree Celsius. 

\begin{table}[h!]
    \centering
    \begin{tabular}{p{1cm} r r}
    \hline
    \textbf{Constant} & \textbf{Original parameters} & \textbf{Calibrated parameters}\\
    \hline
     $b$ & 1396.6000 & 1425.0000\\
     $f$ & 1.0000 & 1.0000\\
     $O$ & -6303.0000 & -9550.0000\\
     $r_1$ & 14911.1846 & 14911.1850\\
     $r_2$ & 0.0108 & 0.0120\\
     \hline
     \textbf{Position} & \textbf{RMSE} & \textbf{MBE}\\
     & \textbf{(in $^o$C)} & \textbf{(in $^o$C)}\\
     \hline
     a & 1.24 & -0.21\\
     b &  1.69 & 0.91\\
     c &  1.00 & 0.55 \\
     \hline
    \end{tabular}
    \caption{Parameters resulting from the calibration of the FLIR A300 camera (see \ref{sec:appendix_B}).}
    \label{tab:calibration_a300_thermal_camera}
\end{table}

\begin{figure}[h!]
    \centering
    \includegraphics[width=9.5cm]{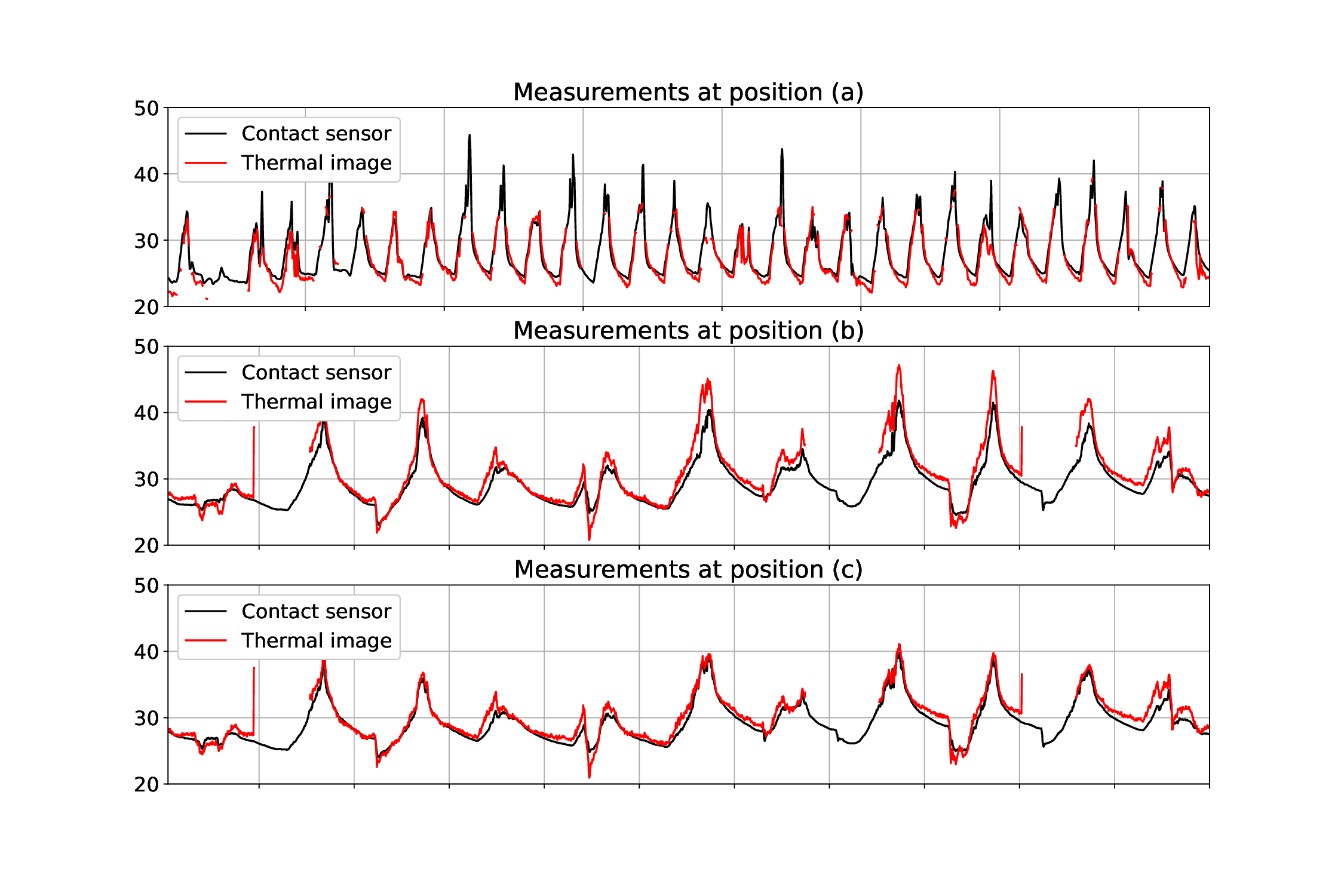}
    \caption{Comparison between estimates and measurements of the surface temperature after calibration of the A300 infrared camera when installed on the observatory.}
    \label{fig:comparison_measurements_estimates}
\end{figure}

Figure \ref{fig:comparison_measurements_estimates} shows a comparison of the surface temperature as assessed from the observatory after calibration and as measured by contact surface sensors at Positions A-C. At Position A, most underestimates seems to occur at night, which might be due to the fact that the actual thermal emissivity of  glass is slightly different from the one considered when estimating the surface temperature. In contrast, during the day, the surface temperature assessed from the observatory looks to be higher than this measured by contact sensors, in particular at Position B. A possible explanation is that the surface temperature measured by a contact sensor can be lower than the actual one as it needs to be fixed with tape on a wall. The tape may generate a small shadow effect around the temperature probe; and thus, create a cooler zone than its surroundings that are directly exposed to the sun. 

\subsection{Heat fluxes of building fa\c{c}ades, vegetation, and traffic}

\begin{figure}[h!]
    \centering
    \includegraphics[width=9.5cm]{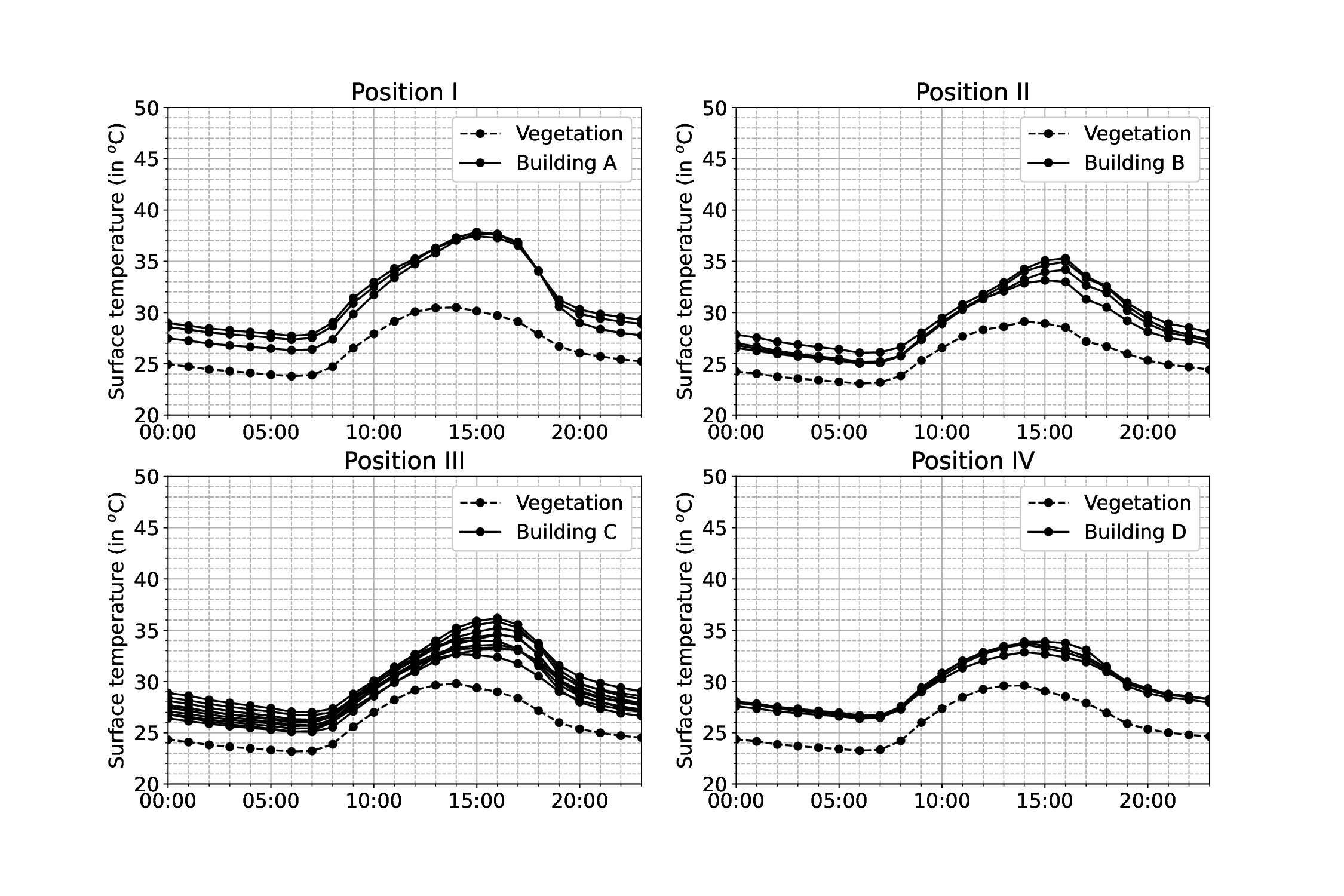}
    \caption{Hourly average surface temperature of building fa\c{c}ades A, B, C, and D in comparison to vegetation at Positions I, II, III, and IV between November 2021 and March 2022.}
    \label{fig:surface_temperature_building_facades_green_areas}
\end{figure}

After calibrating thermal images collected by the observatory, the surface temperature of building fa\c{c}ades and vegetation were assessed using Equation \ref{eq:surface_temperature}. From results shown in Figure  \ref{fig:surface_temperature_building_facades_green_areas}, it appears that fa\c{c}ades of buildings A, B, C, and D are hotter than vegetation over a typical day between November 2021 and March 2022. Among the fa\c{c}ades, the one of Building A seems to be the hottest, in particular between 2 pm and 5 pm when highly exposed to the sun. A similar peak in the surface temperature can be observed on fa\c{c}ades of Buildings B and C but at a lower magnitude than the fa\c{c}ade of Building A. In contrast, no peak is observed between 2 pm and 5 pm on fa\c{c}ades of Building D.

These results prove that the surface temperature of built-up surfaces can be compared to this of vegetation with a higher temporal resolution than what is usually achieved in the literature. A study like the one conducted by \cite{lazzarini2013temperature} show that the surface temperature of built-up surface and vegetation can be analyzed on a monthly basis using thermal images obtained by satellite. It is thus not possible to use satellite thermal images to determine how cool vegetation is in comparison to built-up surfaces on a hourly basis. As implied by \cite{santamouris2015analyzing}, this limitation becomes a major issue when studying a phenomenon like UHIs whose indicators are commonly calculated using measurements of the surface or air temperature that are collected every hour at least.

\begin{figure}[h!]
    \centering
    \includegraphics[width=9.5cm]{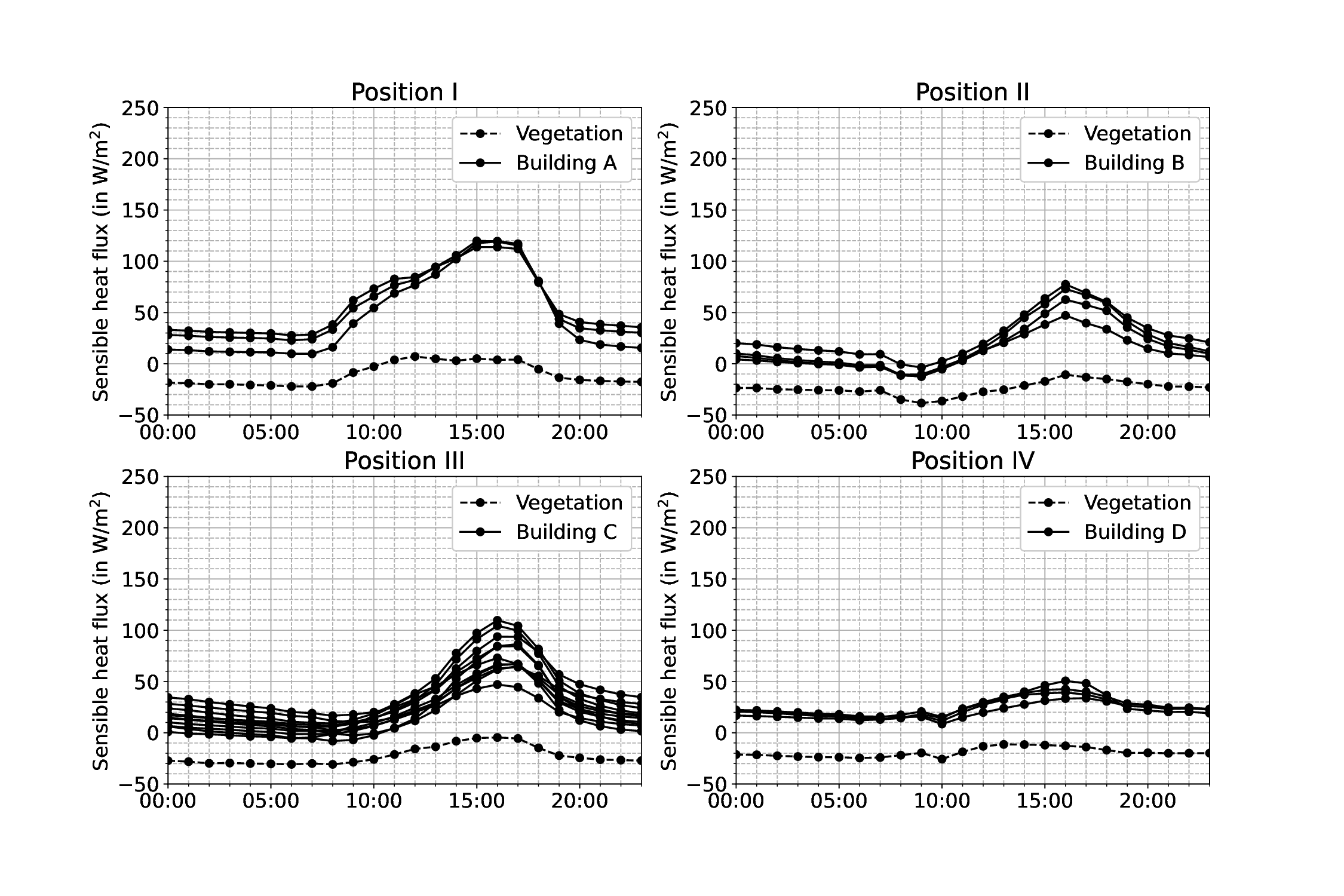}
    \caption{Hourly average convective heat flux of the fa\c{c}ades of Buildings A, B, C, and D in comparison to vegetation at Positions I, II, III, and IV between November 2021 and March 2022.}
    \label{fig:sensible_heat_flux}
\end{figure}

As a consequence of the improved temporal resolution of the surface temperature collected by the observatory in comparison to satellite, Figure \ref{fig:sensible_heat_flux} demonstrates that the convective heat flux of built-up surfaces and vegetation can also be analyzed on a hourly basis. One study in the literature could achieve a similar temporal resolution for analyzing the convective heat flux using thermal images, \cite{hoyano1999analysis}. The study also showed that the magnitude of the convective heat flux from building fa\c{c}ades depends on their orientation. However, thermal images collected by \cite{hoyano1999analysis} did not capture vegetation. It was then not possible to see that the magnitude of the convective heat flux from building fa\c{c}ades is considerably higher than these of vegetation during daytime, as it is observed in Figure \ref{fig:sensible_heat_flux}.

\begin{figure}[h!]
    \centering
    \includegraphics[width=9.5cm]{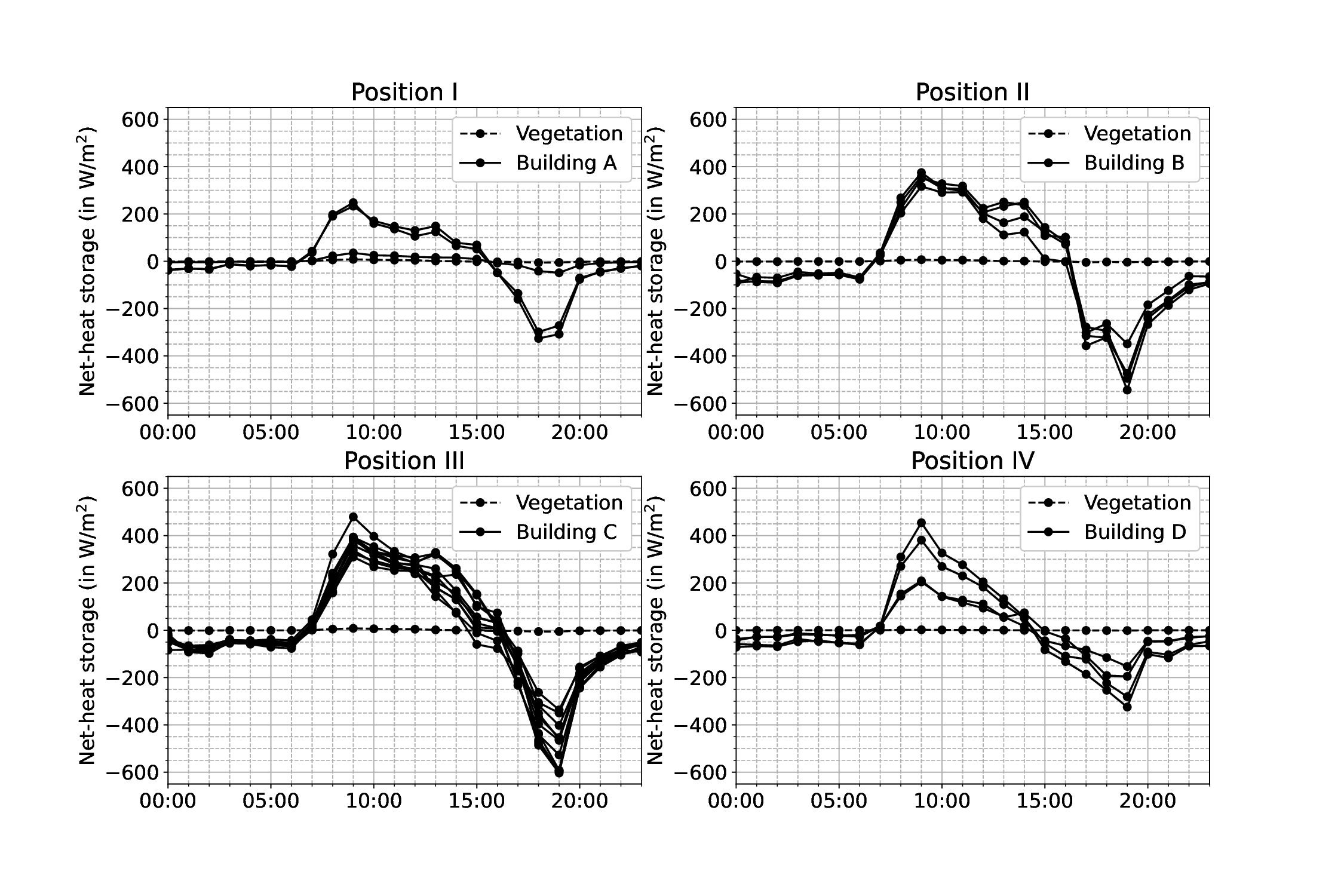}
    \caption{Hourly average net-heat storage of the fa\c{c}ades of Buildings A, B, C, and D in comparison to green areas at Positions I, II, III, and IV between November 2021 and March 2022.}
    \label{fig:net_heat_storage}
\end{figure}

Besides convective heat fluxes, the net-heat storage is another important aspect to observe for understanding the causes of UHIs. As theorized by \cite{oke1981canyon}, UHIs are essentially caused by the heat absorbed by built-up surfaces during the day, which is then released at night. This theory seems to be confirmed by results shown in Figure \ref{fig:net_heat_storage}. In comparison to the vegetation, building fa\c{c}ades accumulate a considerable amount of heat between 6 am and 3 pm. The accumulated heat appears to be then released at a high rate between 4 pm and 8 pm and at a lower one over the night. The high rate of heat loss by building fa\c{c}ades partially coincides with intense sensible heat fluxes observed between 2 pm and 5 pm in Figure \ref{fig:sensible_heat_flux}. Another reason is certainly the increase of the net-longwave radiation from building fa\c{c}ades when their exposure to the Sun decreases.

To affirm that the heat loss between 4 pm and 8 pm is due to an augmentation of the heat emitted by longwave radiation, which is an important component of the sensible heat flux together with the convective heat flux, it would have been required to install a system for measuring the net-all wave radiation in the observatory. Despite this limitation, some assumptions can be made from data collected by weather stations. According to weather stations, the downward shortwave radiation considerably decreases from 2 pm on average over a typical day. From 5 pm, the convective heat flux from building fa\c{c}ades also seems to decrease over time. The important heat loss observed between 4 pm and 8 pm can thus be the result of two phenomena. One hypothesis could be an increase in the heat penetration into buildings by conduction through their fa\c{c}ades. The other hypothesis is then that building fa\c{c}ades lose heat between 4 pm and 8 pm because their radiative heat exchanges with the sky and their surrounding elements as simulated by \cite{miguel2021physically}.

\begin{figure}[h!]
    \centering
    \includegraphics[width=9.5cm]{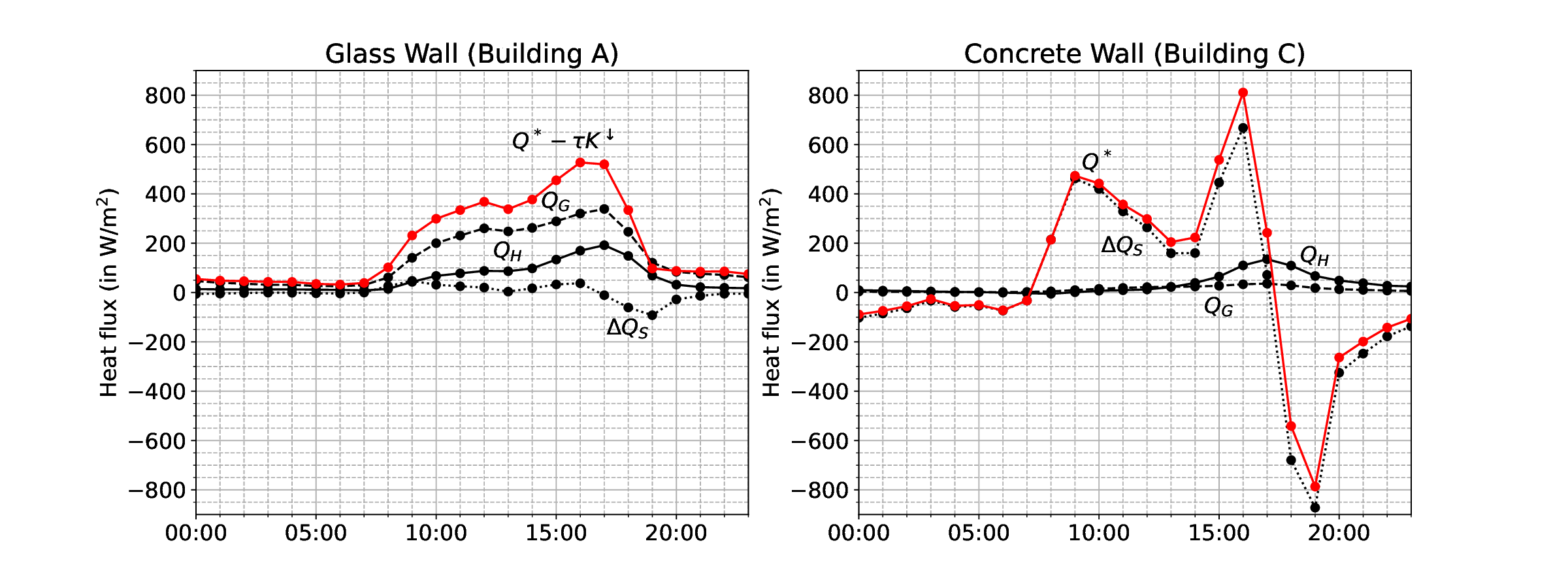}
    \caption{Hourly average heat fluxes of the glass wall (Building A) and the concrete wall (Building C) during sunny days between February 2022 and March 2022.}
    \label{fig:heat_fluxes_builtup_surfaces}
\end{figure}

Figure \ref{fig:heat_fluxes_builtup_surfaces} shows all heat fluxes from the glass wall of Buildings A and from the concrete walls of Building C as assessed from thermal images collected by the observatory, measurements of weather stations, and the inner surface temperature obtained by contact sensors. From this result, the following observations can be made:
\begin{itemize}
    \item The conductive heat flux through glass wall of Building A appears to be relatively high, even at night. It implies that the decrease in the net heat storage between 5pm and 6pm is primarily provoked by convective and conductive heat fluxes.
    \item The net-all wave radiation flux seems to remain positive on the glass wall all over a typical day, which means that more radiations are absorbed by the surface or transmitted into Building A than reflected or re-emitted into the outdoor environment. It explains why highly-glazed or high-rise buildings are more of an indirect contributor of UHIs due to waste heat releases by HVAC systems. 
    \item The conductive heat flux seems to be relatively low through the concrete wall of Building C. There is then a high probability that the decrease in the net-heat storage during the evening is essentially caused by outgoing longwave radiations from the concrete wall.
    \item The net-all wave radiation appears to remain negative on the concrete wall at night. The reason is that the concrete wall certainly re-emits more longwave radiation at night than it absorbs. It can thus be inferred that concrete walls are more of a direct contributor to UHIs because of nighttime outgoing longwave radiation. 
\end{itemize}

Although these observations do not seem to add a considerable knowledge to the contribution of building fa\c{c}ades on UHIs, they have not been illustrated with this level of detail in any study using satellite thermal images or UCMs. On the one hand, satellite thermal images can only capture vertical surfaces, and thus, cannot assess heat fluxes from glass and concrete walls \cite{ngie2014assessment, deilami2018urban}. On the other hand, UCMs often ignore glass walls and do not properly evaluate the solar exposure on concrete walls \cite{ching2013perspective, garuma2018review, jandaghian2020comparing}. The latter limitation is explained from the fact that UCMs assume street canyons to be symmetric.

\begin{figure}[h!]
    \centering
    \includegraphics[width=9.5cm]{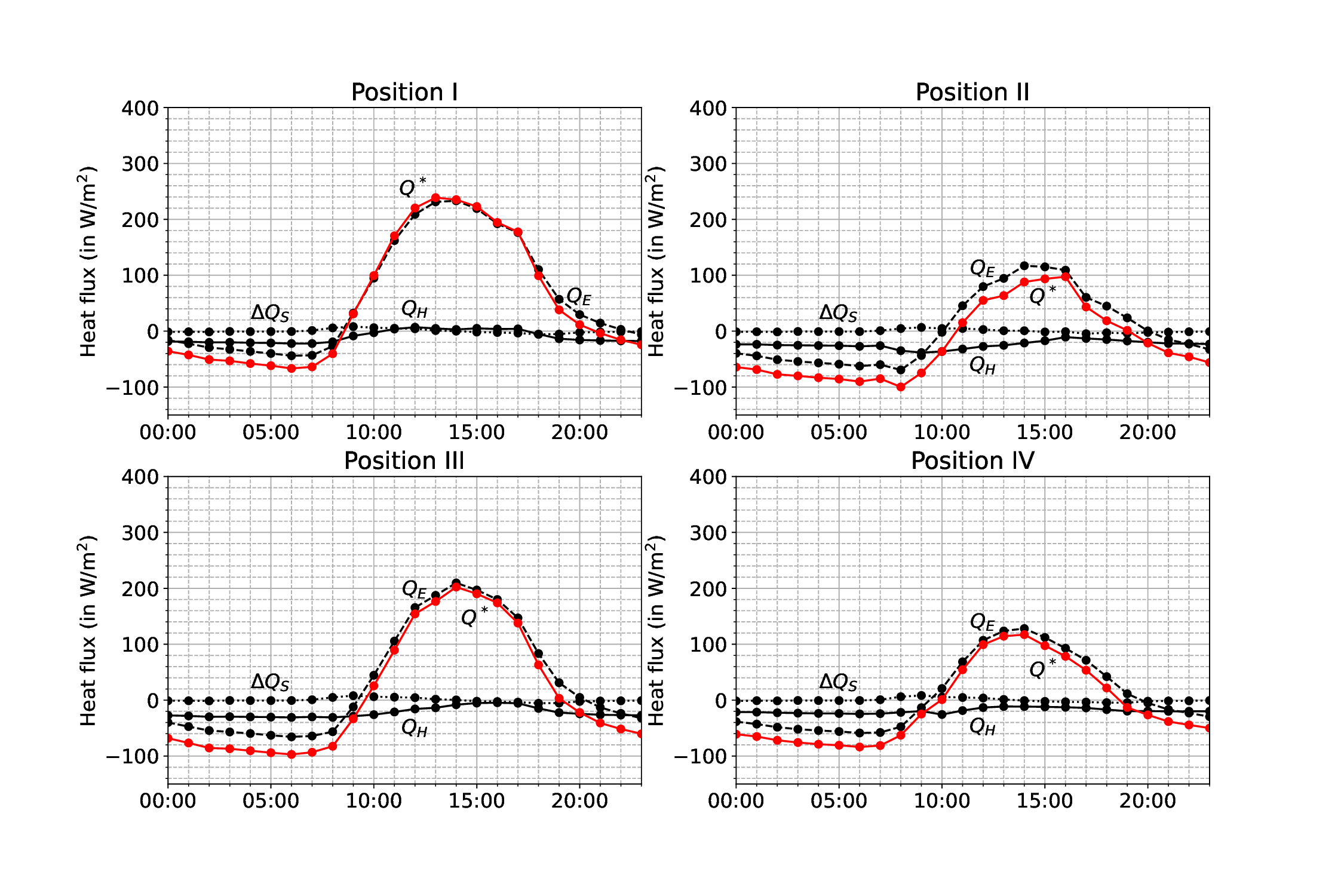}
    \caption{Hourly average heat fluxes of trees at Positions I, II, III, and IV between November 2021 and March 2022.}
    \label{fig:heat_fluxes_trees}
\end{figure}

Similarly to observations made on concrete walls, Figure \ref{fig:heat_fluxes_trees} shows that vegetation appears to emit more longwave radiations than it absorbs at night. However, the heat emitted by longwave radiations looks to be balanced by the convective heat absorbed by the vegetation, which makes the net heat added into the outdoor environment negligible. While the vegetation does not seem to add or extract much heat from the outdoor environment at night, it appears to generate a visible net-cooling effect during the day. The net-cooling effect is largely due to the fact that the vegetation emits little sensible heat by convection and releases a considerable amount of latent heat. This observation implies that vegetation refreshes the outdoor environment by evapotranspiration more than warming it by convection from its surface. At Positions II, III, and IV, it seems that vegetation can even directly absorb heat from the outdoor environment by convection during the day, creating a more significant cooling effect.

Heat fluxes presented in Figure \ref{fig:heat_fluxes_trees} diverge from those usually assessed from methods using UCMs. UCMs, like the one studied by \cite{lee2008vegetated}, often make the big leaf assumption to evaluate heat fluxes of vegetation in a street canyon. In other words, vegetation is modelled as a horizontal surface with a different thermal absorptivity than other built-up surfaces. As a result, the net-heat storage of vegetation is estimated with a slightly lower magnitude than this of built-up surfaces. As reported by \cite{watling2008mechanisms}, and as implied in Figure \ref{fig:heat_fluxes_trees}, plants appear to behave like many living organisms who release sensible and latent heat to thermoregulate their body temperature. The net-heat storage of vegetation is then negligible in comparison to this of built-up surfaces.   

\begin{figure*}[h!]
    \centering
    \includegraphics[width=14cm]{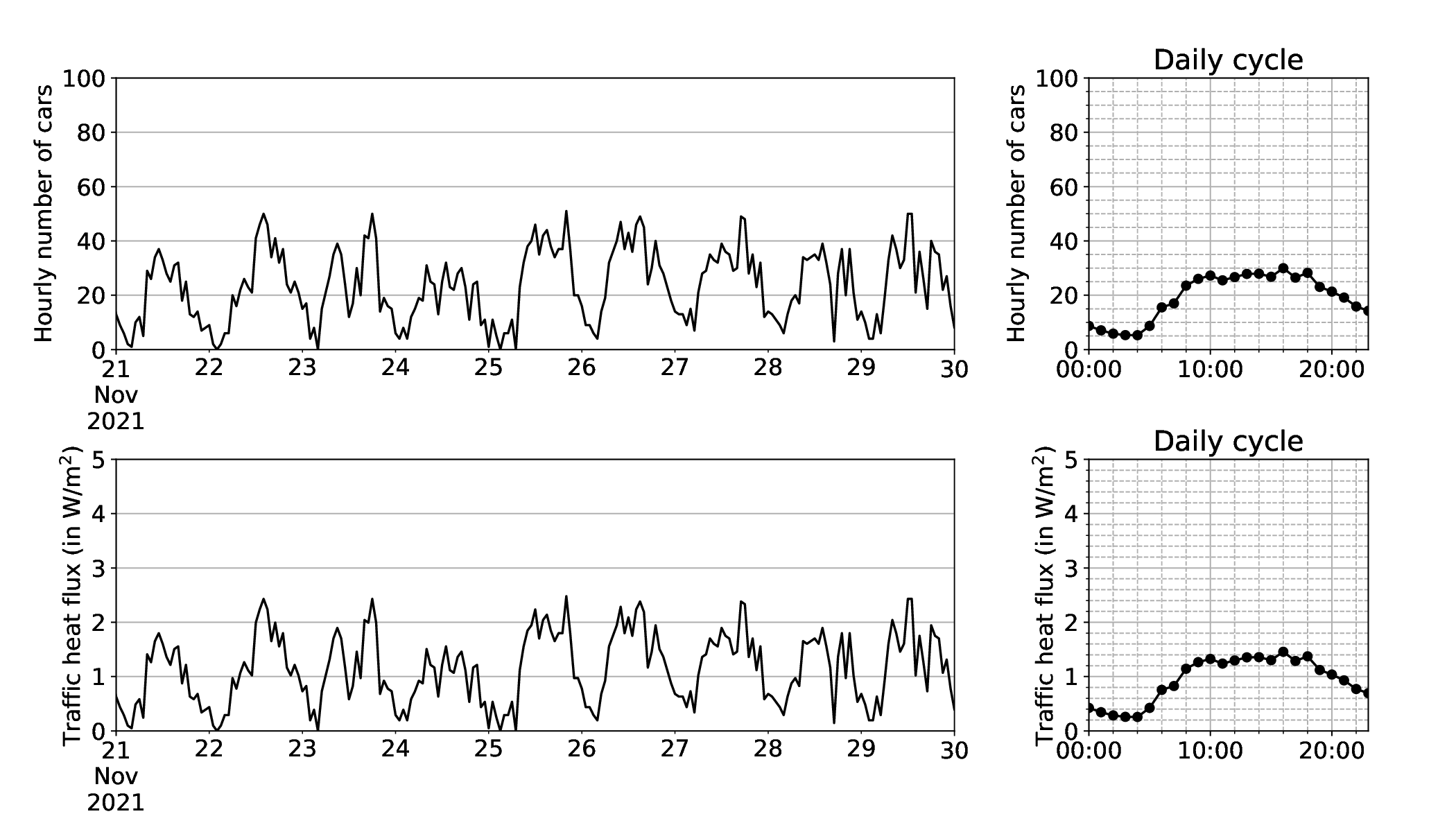}
    \caption{Number of cars and traffic heat flux on a portion of a road at Position IV over the month of November 2021.}
    \label{fig:traffic}
\end{figure*}

Apart from the assessment of heat fluxes from building fa\c{c}ades and vegetation, another major novelty of this study was to estimate the total heat emitted by traffic, both sensible and latent. Figure \ref{fig:traffic} shows the traffic intensity and the total heat flux over the month of November 2021 as estimated using the Equations \ref{eq:number_cars} and \ref{eq:traffic_heat_flux}, respectively. It demonstrates that the traffic is heavier during the day than at night, with two important peaks at 4 pm and 6 pm. The magnitude of the total heat flux from traffic does not go beyond 3 watts per square meter, which is a relatively low value in comparison to other heat fluxes observed in the built environment. The small magnitude of the total heat flux from traffic is justified from the fact that cars were captured by the observatory over a small portion of a road during COVID. The total heat flux from traffic is certainly of much higher magnitude over a larger portion of a road during rush hours in Singapore.

\section{Conclusions}
\label{sec:conclusions}

This study described a method to analyze sensible and latent heat fluxes from building fa\c{c}ades, vegetation, and traffic at the neighborhood scale using thermal images and weather data. While thermal images were obtained from a rooftop observatory, weather data were collected from a network of meteorological stations. Using these two data sources, it was possible to assess heat fluxes from building fa\c{c}ades and green areas within a university campus in Singapore. In addition, the total heat flux from traffic was estimated from a sequence of thermal images.

Before analyzing heat fluxes, a sensitivity analysis was performed on the method used to estimate the surface temperature of elements seen from the observatory. The analysis revealed that the surface temperature assessed from thermal images is more sensitive to some parameters during the day and to others at night. Among the sensitive parameters, there is the outdoor air temperature, which can be measured from weather stations. Other sensitive parameters, however, had to be approximated either from values reported in the literature or from additional instruments. These parameters include the emissivity of elements observed from the infrared camera, their distance from the observatory, and the surface temperature of the window protecting the infrared camera in the protective housing. To improve the accuracy of thermal images collected by an infrared camera, these three variables should be directly measured from devices integrated into the observatory. For example, studies proposed methods to measure the emissivity of different surfaces from satellite \cite{ogawa2002estimation, da2007spectral, ogawa2008estimating}. Another instrument to integrate with the observatory could be a PT100 contact sensor to measure the surface temperature of the window as in the Kipp \& Zonen CNR4 radiometer.

After the sensitivity analysis, some parameters used to estimate the surface temperature were calibrated against measurements in an outdoor environment. With calibrated parameters, it was possible to achieve an RMSE below 2 degrees Celsius and an MBE below $\pm$ 1 degree Celsius between measurements and estimates. The probability that the discrepancy between measurements and estimates is higher than 2.5 degrees Celsius is relatively low. However, this accuracy was achieved from measurements of the surface temperature collected by a few contact sensors installed on opaque surfaces. In the future, it would be recommended to calibrate thermal images with more contact sensors which could also be installed in green areas.

Besides the sensitivity analysis and calibration, the study shows that thermal images collected from an observatory and data obtained from weather stations can be used to observe heat fluxes from built-up surfaces and vegetation with a higher temporal resolution at the neighbourhood scale than using a method relying on satellite data \cite{ngie2014assessment, deilami2018urban}. Heat fluxes can also be analyzed over a longer period of time than using CFD \cite{toparlar2017review, mirzaei2021cfd} and considering urban morphology with a higher fidelity than UCMs \cite{ching2013perspective, garuma2018review, jandaghian2020comparing}. Despite these improvements, thermal images collected from an observatory remain observations, which means that the impact of contributors or mitigators of UHIs cannot be forecasted directly from them. To overcome this current limitation, it would be recommended to use thermal images as an input of data driven models.

Another limitation of the study comes from the fact that heat fluxes were analyzed using thermal images collected by only one observatory. Even though a pan/tilt unit was installed on the observatory to collect thermal images at different positions, some contributors to UHIs like pavement or other land cover materials could not be captured. The impact of surrounding elements to buildings on their thermal behaviour can also hardly be captured from a single point of observation. In the future, it would then be recommended to assess heat fluxes using thermal images collected by multiple infrared systems at several locations as highlighted by the review published by \cite{martin2022infrared}.    

Apart from building fa\c{c}ades and vegetation, the study proves that more dynamic processes in an urban area like traffic can be observed from thermal images collected from an observatory. From the number of cars detected from thermal images, it was possible to assess heat releases from traffic over a portion of the road. This procedure could be applied to several positions to have a better understanding of the impact of traffic on UHIs at the city-scale. In addition to that, it would be interesting to explore whether heat releases from HVAC systems, another important source of anthropogenic heat, could be assessed from thermal images collected by an observatory.

\section*{Acknowledgement}

This research is supported by the National Research Foundation and the Prime Minister's Office of Singapore under its Campus for Research Excellence and Technological Enterprise (CREATE) program. It was funded through a grant to the Berkeley Education Alliance for Research in Singapore (BEARS) for the Singapore-Berkeley Building Efficiency and Sustainability in the Tropics (SinBerBEST) program. BEARS was established by the University of California, Berkeley, as a center for intellectual excellence in research and education in Singapore.

The University Campus Infrastructure of the National University of Singapore also provided great assistance to this research by sharing data obtained during the Virtual Campus project.

\appendix
\section{Calculation of heat balance components}
\label{sec:appendix_A}

This section explains how some heat fluxes are usually expressed or calculated in urban climate studies. It includes the mathematical formulation of the net-all wave radiation ($Q^*$), the sensible heat flux by convection ($Q_H$), the latent heat flux ($Q_E$), the net-heat stored by a surface ($\Delta Q_S$), and the conductive heat flux ($Q_G$). Mathematically, $Q^*$ is defined from the shortwave radiation emitted by the sun on the surface ($K^\downarrow$), the longwave radiation emitted by surroundings on the surface ($L^\downarrow$), the portion of shortwave radiation reflected by the surface ($K^\uparrow$), and the longwave radiation emitted by the surface ($L^\uparrow$) as:
\begin{equation}
    Q^* = K^\downarrow + K^\uparrow + L^\downarrow + L^\uparrow
    \label{eq:net_all_wave_radiation}
\end{equation}
The net-longwave radiation  $L^* = L^\downarrow + L^\uparrow$ can be directly calculated using the following formula:
\begin{equation}
    L^* = \varepsilon \sigma F_{sky} \left( T^4_{sky} - T^4_S \right) + \varepsilon \sigma \sum_{S' \in \mathcal{N}_S}{F_{S \rightarrow S'} \left(T^4_{S'} - T^4_S \right)}
    \label{eq:net_longwave_radiation}
\end{equation}
where $T_{sky}$ is the sky temperature, $\varepsilon$ the emissivity of the surface, $F_{sky}$ its sky view factor, $\mathcal{N}_S$ the set of its surrounding surfaces in the urban area, $F_{S \rightarrow S'}$ its view factor with one of the surrounding surface, and $T_{S'}$ the temperature of one of its surrounding surface. $Q_H$ can be calculated using the following formula:
\begin{equation}
    Q_H = h \left(T_S - T_{air}\right)
    \label{eq:sensible_heat_flux}
\end{equation}
where $h$ is the convective heat transfer coefficient between the surface and the outdoor air, $T_S$ the temperature of a solid surface, and $T_{air}$ the outdoor air temperature. There are various methods to evaluate $h$, but one of the simplest is to use an empirical relation with respect to wind speed $W_s$, that is:
\begin{equation}
    h = b + a \cdot W_s^c
    \label{eq:convective_heat_transfer_coefficient}
\end{equation}
For instance, \cite{masson2000physically} considered that $a = 4.2$, $b = 11.8$, and $c = 1$ for all built-up surfaces inside the urban canopy. Similarly to $Q_H$, $\Delta Q_S$ can be directly calculated from $T_S$ as:
\begin{equation}
    \Delta Q_S = \Delta x \rho c_p \frac{\Delta T_S}{\Delta t}
\end{equation}
where $\Delta T_S / \Delta t$ is the rate of temperature change of the surface, $\Delta x$ its thickness, $\rho$ its density, and $c_p$ its specific heat capacity. $Q_G$ can be obtained using the Fourier first law of heat conduction, that is:
\begin{equation}
    Q_G = \frac{k}{\Delta x} \left(T_S - T_{S_{in}}\right)
\end{equation}
where $k$ is the thermal conductivity of the surface, and $T_{S_{in}}$ the inner layer temperature. 

Using the formulation of \cite{chrysoulakis2018urban}, $Q_E$ can be estimated from the surface temperature of vegetation ($T_S$) as:
\begin{equation}
    Q_E = \frac{\rho c_p}{\gamma} \left(\frac{e^*_S(T_S) - e_{air}}{r_a + r_s}\right)
    \label{eq:latent_heat_flux}
\end{equation}
where $\rho$ is the density of the outdoor air, $c_p$ its specific heat, $\gamma$ a constant (= 0.67 hPa/K), $e^*_S(T_S)$ the saturated vapour pressure at temperature $T_S$ of the vegetation, $e_{air}$ the water pressure of the outdoor air, $r_a$ the aerodynamic resistance between the vegetation and the outdoor air, and $r_s$ the stomatal resistance. Instead of using the log-law, $r_a$ can be expressed from $h$ as:
\begin{equation}
    r_a = \rho c_p / h
\end{equation}
The net-all wave radiation ($Q^*$) and the sensible heat flux by convection ($Q_H$) of the vegetation can be calculated as in Equations \ref{eq:net_all_wave_radiation} and \ref{eq:sensible_heat_flux}, respectively. However, the net-heat stored by the vegetation ($\Delta Q_S$) needs to be calculated as:
\begin{equation}
    \Delta Q_S =  c_{veg} \frac{\Delta T_S}{\Delta t}
    \label{eq:heat_storage_vegetation}
\end{equation}
where $c_{veg}$ is the heat capacitance of the vegetation. According to \cite{lee2008vegetated}, $c_{veg}$ can be expressed as:
\begin{equation}
    c_{veg} = 4186 \cdot LAI
\end{equation}
where $LAI$ is the leaf area index. According to \cite{lee2008vegetated}, $c_{veg}$ can be expressed as:
\begin{equation}
    c_{veg} = 4186 \cdot LAI
\end{equation}

\section{Assessment of surface temperature using thermal images collected by the observatory}
\label{sec:appendix_B}

Regardless of whether thermal images are saved in JPEG or FFF, they contain a header and an array $U^r_{ij}$ corresponding to the output voltage measured by the infrared receptor at position $ij$ of the thermal image. $U^r_{ij}$ can be converted into longwave radiations between 7.5 and 13 micrometers ($L^r_{ij}$) using the following expression:
\begin{equation}
    U^r_{ij} = c L^r_{ij}
\label{eqn:voltage_longwave_radition}
\end{equation}
where $c$ is a constant. The FLIR A300 (9Hz) thermal camera installed in the observatory was calibrated so that the surface temperature of a target element measured by the infrared receptor ($T^r_{ij}$) satisfies the following relation:
\begin{equation}
    T^r_{ij} = g(U^r_{ij}) = b \ln\left[\frac{r_1}{r_2 \left(U_{ij} + O\right)} + f\right]^{-1}
\label{eqn:surface_temperature_from_output_voltage}
\end{equation}
where $b$, $r_1$, $r_2$, $O$, and $f$ are calibrated parameters. It is important to note that these parameters are obtained in a controlled environment where the true surface temperature of a target element at position $ij$ of the thermal image ($T_{ij}$) is almost equal to $T^r_{ij}$. In an outdoor environment, however, a significant discrepancy can be observed between $T_{ij}$ and $T^r_{ij}$. To minimise the discrepancy $T_{ij}$ and $T^r_{ij}$ in an outdoor environment, parameters in Equation \ref{eqn:surface_temperature_from_output_voltage} usually need to be re-calibrated as explained in Section \ref{sec:calibration}.  

Apart from parameters to be calibrated, several aspects had to be considered when evaluating $T_{ij}$ from the rooftop observatory. The first aspect is that buildings, streets, and trees observed from the observatory are exposed to longwave radiations, which mainly come from the skydome. The longwave radiation from the sky ($L^{sky}$) is then reflected into the air in parallel to the longwave radiation emitted by the element observed from the observatory ($L$). Both $L^{sky}$ and $L$ travel over the atmosphere before reaching the observatory. When reaching the observatory, the longwave radiation emitted by the atmosphere ($L^{atm}$), together with $L^{sky}$ and $L$, are transmitted through the window of the housing before being captured by the infrared receptor. The infrared receptor also receives the longwave radiation from the window ($L^{win}$) in addition to others. 

Considering the relation expressed in Equation \ref{eqn:voltage_longwave_radition} and the aforementioned aspects, the camera output voltage corresponding to $L_{ij}$ ($U_{ij}$) can be formulated as:
\begin{equation}
    U_{ij} = \frac{1}{\varepsilon_{ij}\tau^{atm}_{ij} \tau^{win}} U^r_{ij} - \frac{1 - \varepsilon_{ij}}{\varepsilon_{ij}} U^{sky} - \frac{1 - \tau^{atm}_{ij}}{\varepsilon_{ij}\tau^{atm}_{ij}} U^{atm} - \frac{1 - \tau^{win}}{\varepsilon_{ij}\tau^{atm}_{ij}\tau^{win}} U^{win}
\label{eqn:model_infared_camera}
\end{equation}
where $\varepsilon^{m}_{ij}$ is the thermal emissivity an element at position $ij$ in the thermal image, $\tau^{atm}_{ij}$ the transmissivity of the atmosphere between the element and the rooftop observatory, and $\tau^{win}$ the transmissivity of the window. Knowing $U_{ij}$, $T_{ij}$ can be assessed using Equation \ref{eqn:surface_temperature_from_output_voltage}. While $\varepsilon_{ij}$ and $\tau^{win}$ can be estimated from material properties of the element and the window, respectively, $\tau^{atm}_{ij}$ needs to be calculated from data collected by a weather stations as:
\begin{equation}
    \tau^{atm}_{ij} = x \cdot \exp\left[- \sqrt{d_{ij}} \left( \alpha_1 + \beta_1 \sqrt{\omega}  \right) \right] + (1 - x) \cdot \exp\left[- \sqrt{d_{ij}} \left( \alpha_2 + \beta_2 \sqrt{\omega}  \right) \right]
\end{equation}
where $d_{ij}$ is the distance between an element and the observatory at position $ij$ of the thermal image, $\omega$ is the water vapour content in the atmosphere, $x$, $\alpha_1$, $\beta_1$, $\alpha_2$, and $\beta_2$ empirical coefficients. $\omega$ can be estimated from the air temperature ($T_{air}$) and relative humidity ($\phi$) measured from the weather station as:
\begin{equation}
    \omega = \phi \cdot \exp\left[c_0 + c_1 T_{air} + c_2 T^2_{air} + c_3 T^3_{air} \right]
\end{equation}
where $c_0$, $c_1$, $c_2$, and $c_3$ is another set of empirical coefficients described in \cite{waldemar2015modeling}.

\bibliographystyle{unsrt}
\bibliography{mybib}

\end{document}